\documentclass[twocolumn]{aa}
\usepackage{graphicx,epsf,psfig,natbib}
\usepackage{amsmath,amssymb,rotating}
\usepackage{txfonts}
\begin{document}
   \title{An unbiased search for the signatures of protostars in the 
           $\rho$~Ophiuchi Molecular Cloud
         \thanks{Based on observations collected at the 
                 European Southern Observatory, Chile, 
                 proposal number 69.C-0639}}
\subtitle{II. Millimetre continuum observations}
\titlerunning{The signatures of protostars in $\rho$~Ophiuchus II.\@ Millimetre continuum observations}

   \author{T. Stanke
          \inst{1,2}
          \and
          M.D. Smith\inst{3,5}
          \and
          R. Gredel\inst{4}
          \and
          T. Khanzadyan\inst{4}
          }

   \offprints{T. Stanke}

   \institute{Max-Planck-Institut f\"ur Radioastronomie Bonn,
              Auf dem H\"ugel 69, D-53121 Bonn, Germany
     \and
       Institute for Astronomy, University of Hawai'i, 2680 Woodlawn Drive,
       Honolulu, Hawai'i 96822, USA\\
       \email{stanke@ifa.hawaii.edu}
     \and
       Armagh Observatory, College Hill, Armagh BT61 9DG, Northern Ireland,
       UK\\
       \email{mds@arm.ac.uk}
     \and
       Max-Planck-Institut f\"ur Astronomie, K\"onigsstuhl 17,
       69117 Heidelberg, Germany\\
       \email{gredel@caha.es, khtig@mpia-hd.mpg.de}
     \and
       Centre for Astrophysics and Planetary Science, School of Physical
       Sciences, University of Kent CT2 7NR
             }

   \date{Received ; accepted }

   \abstract{
   The dense cores which conceive and cradle young stars can be explored
   through continuum emission from associated dust grains. We have performed
   a wide field survey for dust sources at 1.2\,millimetres in the 
   $\rho$~Ophiuchi molecular cloud, covering more than 1 square degree in an
   unbiased fashion. We detect a number of previously unknown sources, ranging
   from extended cores over compact, starless cores to envelopes surrounding
   young stellar objects of Class~0, Class~I, and Class~II type. We analyse
   the mass distribution, spatial distribution and the potential equilibrium
   of the cores. For the inner regions, the survey results are consistent with
   the findings of previous narrower surveys. The core mass function resembles
   the stellar initial mass function, with the core mass function shifted by
   a factor of two to higher masses (for the chosen opacity and temperature).
   In addition, we find no statistical variation in the core mass function
   between the crowded inner regions and those in more isolated fields except
   for the absence of the most massive cores in the extended cloud. The inner
   region contains compacter cores. This is interpreted as due to a medium of
   higher mean pressure although strong pressure variations are evident in
   each region. The cores display a hierarchical spatial distribution with
   no preferred separation scale length. However, the frequency distribution
   of nearest neighbours displays two peaks, one of which at 5000~AU can be
   the result of core fragmentation. The orientations of the major axes of
   cores are consistent with an isotropic distribution. In contrast, the
   relative orientations of core pairs are preferentially in the NW-SE
   direction on all separation scales. These results are consistent with
   core production and evolution in a turbulent environment.   
   Finally, we report the discovery of a new, low-mass Class~0 object
   candidate and its CO outflow.

   \keywords{ISM -- Stars: formation -- ISM: clouds -- 
   ISM: individual objects: Rho Ophiuchi -- ISM: structure}
   }

   \maketitle
%

\section{Introduction
\label{chap_intro}}

As one of the nearest star-forming regions, the $\rho$~Ophiuchi 
molecular cloud complex has been the target of numerous investigations.
A prime focus has been the densest part of the \object{L\,1688} cloud which harbours a
large number of young stellar objects, studied at near- to mid-infrared
wavelengths \citep{1975ApJ...197...77V,1978ApJ...224..453E,
1989ApJ...340..823W,1989ApJ...346L..93B,1993ApJ...416..185C,
1999ApJ...525..440L,2001ApJ...551..357W,2001A&A...372..173B,
2002ApJ...566..993A} and at (sub)millimetre wavelengths 
\citep{1993ApJ...406..122A,1994ApJ...420..837A,1998A&A...330..549N,
1999ApJ...513L.139W,2000ApJ...528..817T}.
To investigate how these stars are conceived, we wish to to relate their
properties to those of the embedding cloud of molecular gas and dust which
both nurture and obscure the stars. However, technical capabilities restrict
the field over which area-covering surveys can be undertaken, bearing the
danger of picking peculiar objects prevalent in the ``more interesting''
parts, thus biasing the results obtained from statistical investigations.
In order to further exclude such selection effects, we have
performed an unbiased survey of a wide area around the L\,1688 molecular cloud
for dust continuum sources, including the dense molecular cores as well
as the areas which apparently do not harbour dense molecular material. 

The entire star formation complex extends over several degrees on the sky
\citep{1989ApJ...338..902L}, containing a few major clouds.
The L\,1688 cloud is situated about
one degree south of the \object{$\rho$~Ophiuchus star} itself. This `main cloud' of 
$\rho$~Oph covers an area of roughly 480~arcmin$^2$ and has been dissected
into about a dozen 
cloud components \citep{1992A&A...265..743M}. This area was surveyed 
at 1.3~mm by \citet[][hereafter MAN98]{1998A&A...336..150M}, who uncovered
62 starless cores and 41 circumstellar structures. These observations had a
resolution of just 11~arcsec, corresponding to 1,400~AU. The inferred
distribution of masses of these cores was found to be comparable to the
{\em stellar} initial mass function, suggesting that stellar masses are
determined at conception. Subsequently, the results of a larger, somewhat more 
sensitive survey at 0.85~mm were presented by \citet{2000ApJ...545..327J}
(hereafter J00). This survey covered $\sim$ 700~arcmin$^2$ with a resolution
of 14~arcsec, identifying 55 cores. 
This survey was recently followed up by a much more extensive (4 square
degrees) but significantly more shallow survey by \citet{2004ApJ...611L..45J}
(hereafter J04), mainly extending to the north/north-west of the main cloud.

In the present 1.2~mm survey, we cover an area of 4,600~arcmin$^2$.
This constitutes a significant extension to the above works although the
resolution is $\sim$24\arcsec. Our immediate goals are to (i) derive the core
masses (\S \ref{chap_res}), (ii) estimate the mass relative to a critical 
Bonnor-Ebert sphere or Jeans mass by including the core sizes 
(\S \ref{chap_equi}), (iii) quantify the spatial distribution and orientations
(\S \ref{chap_distrib} \& \S \ref{chap_orient}), and (iv) quantify the mass
distribution (\S \ref{chap_massdist}). In addition, we report the discovery
of a new low-mass candidate Class~0 source and its CO outflow 
(\S \ref{chap_mms126}).
A companion paper \citep{2004A&A...426..171K} presents the results of a
near-infrared imaging search for protostellar H$_2$ outflows over a large
section of the present survey area. Finally, the core, outflow and
protostellar properties will be related in detail in a forthcoming paper.

Quoted distances to the $\rho$~Ophiuchi cloud lie in the range 
$D =$ 125 -- 165\,pc, as recently summarised by \citet{2004AJ....127.1029R}.
\citet{1989A&A...216...44D} analysed photometry of the stars in the nearby
Sco~OB2 association. It was shown that the Ophiuchus dark clouds are on the
near side of \object{Upper Scorpius}, at $\sim$ 125\,pc. Hipparcos data confirm the
distance to Upper Scorpius and yield a distance of $\sim$\,128\,pc to the 
quadruple $\rho$ Oph system  \citep{1999AJ....117..354D}. Here, we shall
follow the suggestion by \citet{2004AJ....127.1029R} and take the 
weighted average distance of 130\,pc.

In the following, we will designate all well-defined low mass
structures as `cores' (with mass below about $50\,$M$_\odot$)
and apply the term `clump' to more
massive structures defined through low resolution CO mapping. A clump
usually corresponds spatially to a cluster of cores.  

\section{Observations       
\label{chap_obs}}           
\begin{figure*}
  \centering
  \vspace*{1cm}
  {\bf available as jpg: 0511093.1331f01.jpg}\\
  {\bf full postscript version available at http://www.ifa.hawaii.edu//users/stanke/preprints.html}
  \vspace*{1cm}
  \caption{SIMBA 1.2~mm continuum mosaic image of the $\rho$--Ophiuchi cloud.
  Intensity scaling is in mJy/beam (linear). Prominent clumps are marked as
  A, B1/1, C, D, E, and F. Dashed lines mark some filamentary structures
  east of the major clumps, with three more compact sources (MMS038, MMS051,
  and MMS126) within them also labeled. MMS126 is identified as a new, low-mass
  Class~0 object candidate in Sect.\ \ref{chap_mms126}.}
  \label{fig_mosaic}
\end{figure*}

The observations were carried out using the SIMBA bolometer array at the SEST
telescope on La Silla/Chile during an observing run lasting from
2002 July 7 to 12. SIMBA is a 37-channel bolometer array, operating at 1.2~mm,
and provides a HPBW of $\sim$24\arcsec{} (as measured on maps of Uranus).
The weather was good and fairly 
stable with zenith opacities ranging between 0.17 and $\sim$\,0.3.
In total, 78 maps in the fast mapping mode were taken (i.e., no wobbler was
used). Typical map sizes were 1200\arcsec{}$\times$800\arcsec{} with a
scanning velocity of 80\arcsec{}/sec and 8\arcsec{} steps between subscans.
A number of maps were taken with a scanning velocity of 160\arcsec{}/sec
(size 1800\arcsec{}$\times$800\arcsec{}) in order to improve sensitivity
for extended structures. Pointing and focus were checked regularly;
pointing corrections were typically a few arcsec and always below 10\arcsec{}.
Skydips were taken every 1 to 2 hours, and Uranus
was mapped several times for calibration purposes. 

The data reduction was performed with MOPSI, a software package developed
by Robert Zylka, following standard bolometer data reduction principles.
First, noisy or dead channels were identified and deleted. Then, a low order 
baseline was subtracted from the data and an initial de-spiking was made.
The data direct from the telescope consist of the true signal convolved with
the response of the electronics, hence requiring
deconvolution \citep{2001A&A...379..735R,2002A&A...383.1088W}.
We chose to reconstruct the signal at a specific time using only data from
a given window (typically 6 seconds) around this time, including the data
recorded during the ``turning'' of the telescope between two 
on-map sub-scans (in the standard deconvolution only data taken during
the on-map subscans are used, and the signal is reconstructed from the
entire on-map sub-scan).
This deconvolution was found to provide significantly better results than
the standard deconvolution, for two reasons: the restriction
to a shorter time span reduces noise that goes into the reconstruction
of a given data point, and including the time during the turning of the
telescope ensures a proper reconstruction of the data at the edges of the
map, as sky fluctuations during turning (which are convolved into the
data recorded during the on-map subscans) are properly taken out of the data
recorded during the on-map subscans.
Then, after the deconvolution, the data corresponding to off-map data
taken during the turning of the telescope were deleted.
After further low-order baseline corrections, correction for atmospheric
extinction and correction for gain-elevation effects, correlated sky
brightness variations (sky-noise) were removed. Here, an iterative procedure 
was used: for each iteration, we used the resulting mosaic of the previous 
iteration as an input model for the brightness distribution.
Following another careful de-spiking and baseline subtraction, the data
were then combined into a mosaic, weighting individual maps according
to their noise (1/rms$^2$).
After each iteration, for each sub-map the polygons defining emission-free
regions to be used for baseline fitting were carefully checked and, if 
necessary, adjusted, followed by a check of the degree of the polynomial
used for baseline subtraction.

The noise level in the final mosaic is of the order of 10~mJy/beam
throughout most of the map. Close to the edges, it degrades to 12-15~mJy/beam.
For a point source, assuming a flux-to-mass conversion as outlined
below, this corresponds to a 3$\sigma$ detection limit of the order of
0.014~M$_\odot$. In terms of column density, the 1$\sigma$ noise level
corresponds to about 1.3$\times 10^{21}$cm$^{-2}$. This is slightly better
than the 1.3~mm survey done by MAN98, but somewhat less sensitive than the J00
850$\mu$m survey; it is significantly more sensitive than the recent
J04 850$\mu$m survey.

Similar to other (sub)millimetre surveys, our new map is sensitive only to
sufficiently small structures (see also J00 for a discussion of this issue).
In the case of SIMBA fast mapping observations,
this is due to two effects: the subtraction of the sky emission removes any
flux due to large scale, more or less uniform emission, and a high-pass filter
applied to the signal from the bolometer will further suppress large-scale
emission features. On top of these technical effects, baseline subtraction may
further contribute; however, baseline subtraction was done very carefully
in order to minimise this. Filtering of low frequencies and residuals from
the deconvolution of the data are responsible for residual negative map
values close to the brightest parts of the mosaic (dipping to 
$\sim$-20~mJy/beam (-2~$\sigma$) in the most negative patches), 
a feature generally found in SIMBA maps.

\section{Data Analysis      
\label{chap_res}}           

\begin{figure*}
  \centering
  \vspace*{1cm}
  {\bf available as jpg: 0511093.1331f02.jpg}\\
  {\bf full postscript version available at http://www.ifa.hawaii.edu//users/stanke/preprints.html}
  \vspace*{1cm}
  \caption{Model of the 1.2~mm continuum emission derived from the
     superposition of models for the 143 sources identified from the SIMBA
     data. Intensity scaling is in mJy/beam (linear).}
  \label{fig_model}
\end{figure*}

\begin{figure*}
  \centering
  \vspace*{1cm}
  {\bf available as jpg: 0511093.1331f03.jpg}\\
  {\bf full postscript version available at http://www.ifa.hawaii.edu//users/stanke/preprints.html}
  \vspace*{1cm}
  \caption{Residual after subtracting the model source distribution (Fig.\
     \ref{fig_model}) from the 1.2~mm mosaic (Fig.\ \ref{fig_mosaic}).
     Intensity scaling is in mJy/beam (linear).}
  \label{fig_residual}
\end{figure*}

%

The final 1.2~mm mosaic is shown in Fig.~\ref{fig_mosaic}. In line with
previous (sub)millimetre maps, a large variety of sources is seen, ranging
from extended cores over more compact features to unresolved point sources.
In those regions which were also covered by the MAN98 and J00 surveys
we generally see a very good agreement between the different data sets
(limited by the poorer angular resolution of our new data).

In the part of the mosaic which had not been previously observed, besides
a number of fainter, extended structures, two more remarkable features
turned up. On the one hand, there is a filamentary structure (dashed lines
in Fig.\ \ref{fig_mosaic}) to the
east of the previously revealed cores Oph-B/C/E/F, containing two
more compact sources (MMS126 and MMS038/MMS051). MMS126 appears to be a newly
discovered, low-mass Class~0 source (see Sect.\ \ref{chap_mms126} for details).

The second remarkable structure is the presence of a bright dust core 
MMS041 at the tip of the cometary
dark cloud \object{L\,1696B}. A similar core was already found in the neighbouring
cloud \object{L\,1696} by MAN98.

\subsection{Source identification 
\label{chap_sourceid}}            
  
The variety in source morphologies makes a detailed source identification
mandatory: e.g., using simple source detection algorithms with a certain
detection threshold will not identify extended structures, which are clearly
present, but might have maximum surface brightnesses of the order of the
1$\sigma$ noise level. To account for this, sources were identified
using a wavelet decomposition similar to the method used by MAN98.

The final map was split into 5 planes using the TRANSFORM/WAVE and
EXTRACT/WAVE tasks provided by the ESO/MIDAS {\em wavelet} context package.
We will refer to the most compact scale as scale 1, and scale 2, 3, 4, and
5 to the next more extended scales.

We then used the 2-D adaption of the {\em clumpfind} algorithm developed by 
\citet{1994ApJ...428..693W}
to identify features in the 5 planes individually. The lower limits for
source identifications were set to 3-4 times the rms fluctuation of 
emission-free parts of the respective planes, with contours spaced by
about 1 rms. The minimum number of pixels required to identify a
sufficiently bright feature as a source was set according to the
characteristic scale of each plane (i.e., to very small for
scale 1, and getting bigger for larger scales).
The thus detected features were then visually inspected to reject
false detections due to residual scanning artifacts (found to be 
restricted to scales 1 and 2) or other misidentifications.

From the list of features identified at the various scales a final 
source list was compiled, including all features identified in scale 1,
and all sources in the more extended scales which did not overlap with
sources identified in the next more compact scale (i.e., for each 
feature we determined the part of the clump area as given by the
{\em clumpfind} routine, in which the intensity was greater than 1/2 the
maximum intensity, and checked if it overlapped with the corresponding
area of all features of the next more compact scale). This resulted in
a list of 139 sources (MMS001 to MMS139).

To this list we added 4 sources,
MMS140 to MMS143, after visual inspection: while MMS141 is just too 
faint to be detectable in the individual planes, MMS140, MMS142, and
MMS143 are small cores adjacent to large bright features, which
make them hard to detect as taking out the extended scales creates
negative haloes around large bright cores.

A full finding chart showing the location of all 143 sources is shown
in Fig.\ \ref{fig_findingchart} (online version only), and closeups of
crowded regions are shown in Figs.\ \ref{fig_findingchart_A},
\ref{fig_findingchart_B}, and \ref{fig_findingchart_CEF} (online version only).

Table \ref{tab_sources} (online version only) lists the 1.2~mm sources found
in our survey in the order they were identified by the wavelet decomposition
plus {\em clumpfind} technique. In addition to the source position, it holds
the integrated source flux and the peak flux determined in various ways,
the major and minor axis and the position angle of the major axis,
cross-identifications to the sources identified by 
\citet{1998A&A...336..150M} and \citet{2000ApJ...545..327J}, 
and a comment on association with known YSOs.
In Section \ref{app_indiv} we give notes on the individual sources.

\subsection{Measuring source properties 
\label{chap_sourceprop}}                
\begin{figure}[h!]
     \centering\includegraphics[width=0.9\linewidth]{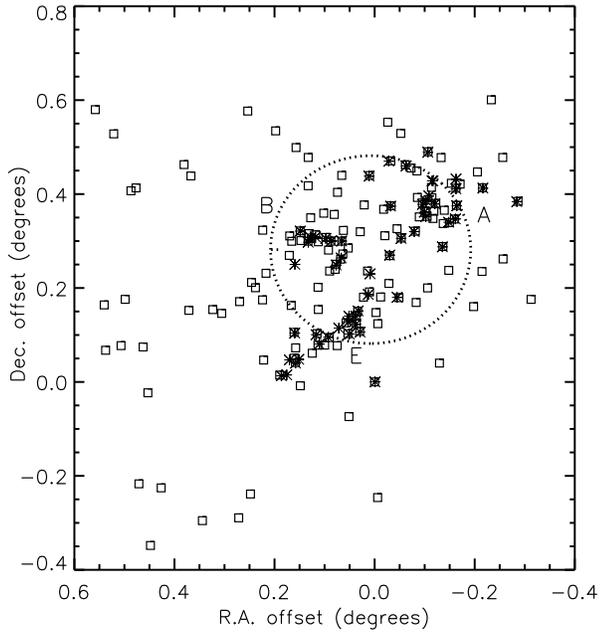}
     \caption[ ]%
      {The locations of all the 143 cores detected here (squares) compared to
      all the cores detected by J00 (asterisks). The
      coordinates are centred on -16:26:58.4, -24:45:36 (2000). A, B, and
      E mark the location of prominent clumps.}
      \label{loc-j}
\end{figure}
\begin{figure}
  \centering\includegraphics[width=0.9\linewidth]{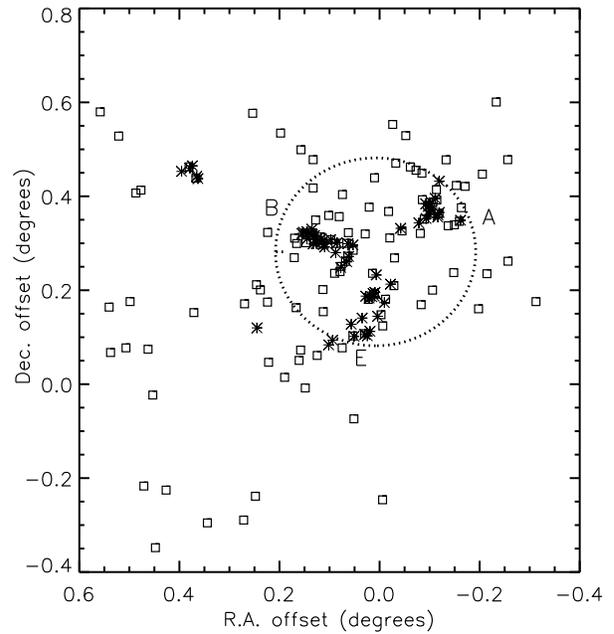}
  \caption[ ]%
    {The locations of the 111 {\em starless} cores detected here (squares)
    compared to the the {\em starless} cores detected by
    MAN98 (asterisks). The circle
    is that employed to separate inner cores from the outer regions.}
  \label{loc-motte}
\end{figure}
  
We then performed a detailed modelling of the overall
brightness distribution using Gaussian sources, the result of which
is shown in Fig.\ \ref{fig_model} (Fig.\ \ref{fig_residual} shows the residual,
after subtracting the model source distribution from the mosaic). In some
cases, more than one Gaussian component centered on the source position had
to be assumed in order to properly reproduce the shape of
the sources (e.g., a bright source with a radial intensity distribution
falling off as a power law will not be satisfactorily fitted by a single
Gaussian).

The sources we detect within the inner core-crowded region are generally
coincident with those found in previous surveys. Figs.~\ref{loc-j}
and \ref{loc-motte} show that this is true for separate samples of all cores
and starless cores. In particular, we recover all the cores from J00, despite 
the different observing wavelength.
This confirms the reliability of our core-finding
technique.

The fluxes of the sources were determined by first subtracting the model for
all other sources from the mosaic, and then integrating over a certain area
around the source within some given radius or (for more complex sources) within
a certain polygon around the centre of the source. Similarly, sizes and
position angles for each source were determined by (single) Gaussian fits to
the mosaic with the model for all other sources subtracted.

Peak fluxes were calculated in four differing ways and are also
given in Table \ref{tab_sources}: the column labelled as F$_{\rm peak}$
(a) gives the peak
flux obtained from searching the maximum pixel value in an ellipse
(with the minor and major axes as given in the table) around the
peak position; F$_{\rm peak}$ (b) is the same as (a), but
using only the sum of scales 5 down to the scale where the source
was detected first; F$_{\rm peak}$ (c) gives the maximum pixel value in the
same ellipse after re-binning according to the size of the
sources minor axis (i.e., by FWHM$_{\rm min}$/24\arcsec pixels);
F$_{\rm peak}$ (d) gives the maximum pixel value of the model of
the source. As done in deriving the total
source flux, the models of all other sources were subtracted from the
mosaic before determining the respective peak fluxes.

Depending on the source morphology and brightness, one or the other
method can be expected to yield more accurate results. Relatively
bright, compact sources should be best represented by F$_{\rm peak}$ (a),
whereas faint, extended sources should be best measured by F$_{\rm
peak}$ (b) or (c). Generally, the results agree well, apart from an obvious
overestimate for large, low surface-brightness sources using the first
method  and systematic underestimates for compact sources using the
second and third methods.

At this point it might be worth adding a note on the dependency of the
effective mass/column density sensitivity limit on source
size. The surveys of J00 and J04 identified sources setting a
fixed threshold in peak flux, i.e., column density. In contrast, we also
include sources whose surface brightness is nominally lower than,
e.g., 3$\sigma$, but which are still clearly recognised because they
are extended (the MAN98 analysis using a wavelet decomposition might
be comparable in this respect). This introduces a size-dependent
column density limit: for a source which has twice the radius ${\rm R}$, we
can rebin the mosaic by 2$\times$2 pixels, effectively reducing the
noise by a factor of two. I.e., the column density detection limit
scales as ${\rm R}^{-1}$, which implies a mass detection limit scaling as
${\rm R}^1$ (whereas a fixed surface brightness cutoff implies a mass
detection limit scaling as ${\rm R}^2$; always assuming that all sources
have the same temperature and dust properties).
In this sense, we can expect that our method of source identification
yields, for a given mass, a more complete census than applying simply
a column density cutoff, because it also includes the more extended,
hence fainter (in surface brightness), sources.

Our method of source identification yielded significantly more sources
than identified by J00 in areas covered by both surveys, despite the somewhat
poorer angular resolution and slightly lower sensitivity. This is due to two
reasons. First, J00 explicitly filter out more extended emission before
applying {\em clumpfind}, thus removing extended features and lowering the
signal also for more compact features. Second, they limit their identification
to features with a given column density threshold, again biasing against
low surface brightness, extended sources. Detailed comparison of our data with
the J00 map before filtering out the extended emission shows that virtually all
features that we identify are also seen on the J00 map, even if not identified
in their list of objects. We have marked those features as 'n.i.' in the
'J00' column of Tab.\ \ref{tab_sources}.

\section{Core Properties   
\label{chap_equi}}          

\subsection{Masses and column densities}

\begin{figure}[th]
\centerline{\psfig{file=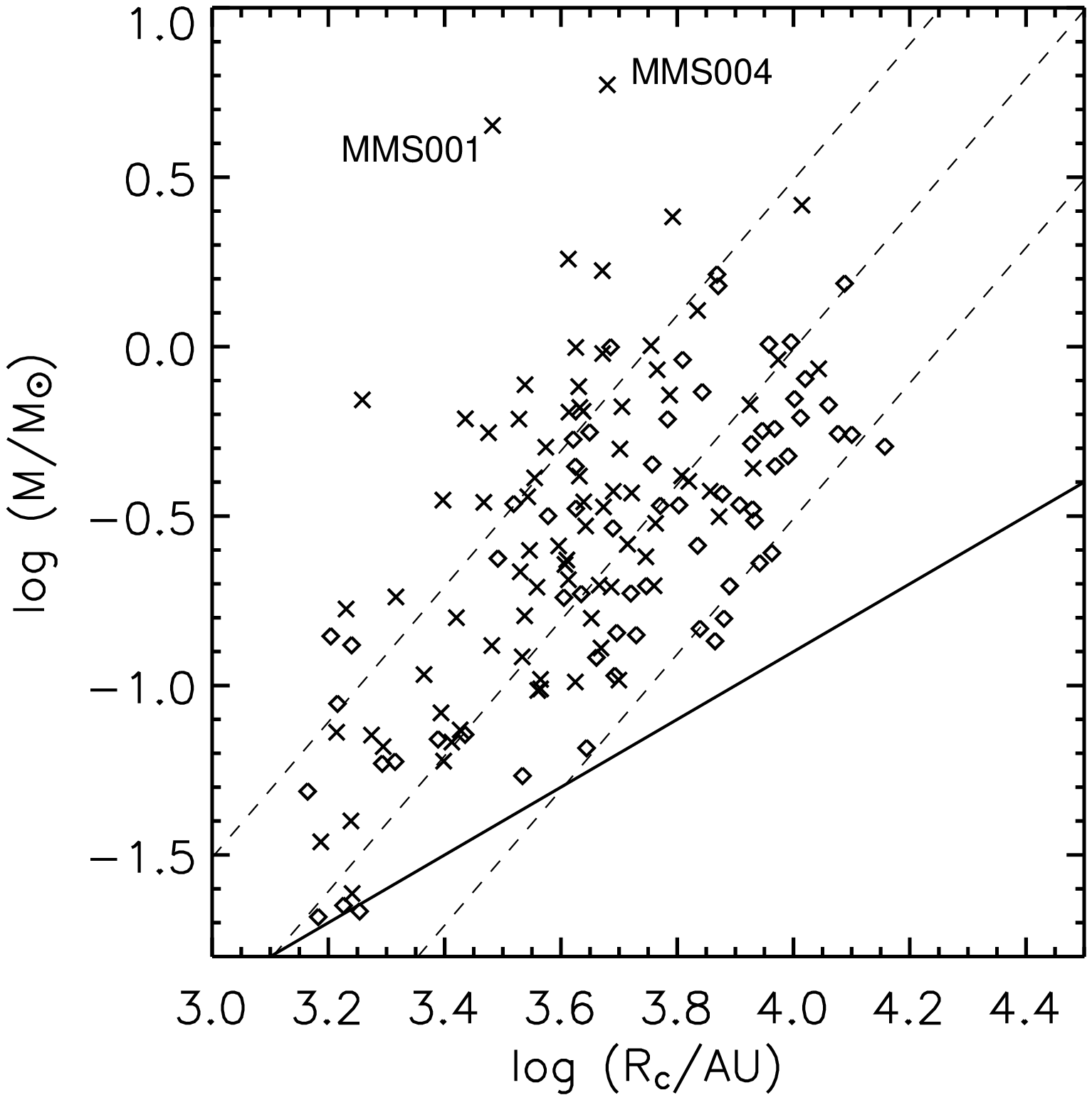,width=8.7cm}}
\centerline{\psfig{file=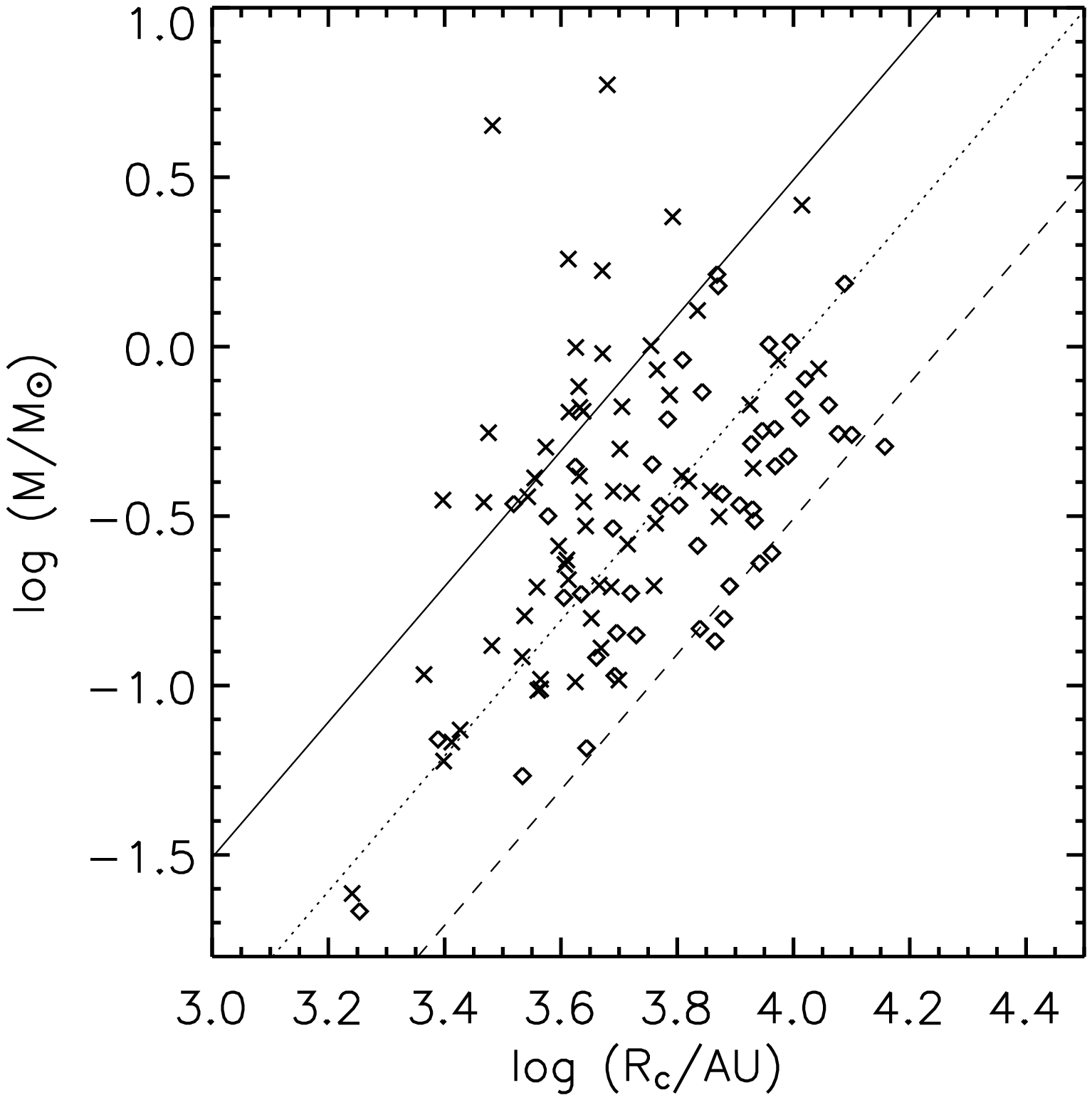,width=8.7cm}}
\vspace*{8pt}
\caption{{\bf Upper Panel}: The mass and radius derived from the observations
   for each core. The crosses correspond to the 79 cores found within the
   central circle of radius 0.2$^\circ$ shown in Fig.~\ref{mf-circle} and
   diamonds to those 64 outside. Lines of constant column density are
   overplotted for reference as dashed lines, with columns of atomic
   hydrogen from $3 \times 10^{21}$ cm$^{-2}$ (right) to  
   $3 \times 10^{22}$ cm$^{-2}$ (left). The solid line marks the detection
   limit.
  {\bf Lower Panel:}
   The masses and radii for just the starless cores, consisting of 62 within
   the circle (crosses) and 49 outside (diamonds).}
\label{massrad}
\end{figure}

\begin{figure}[th]
\centerline{\psfig{file=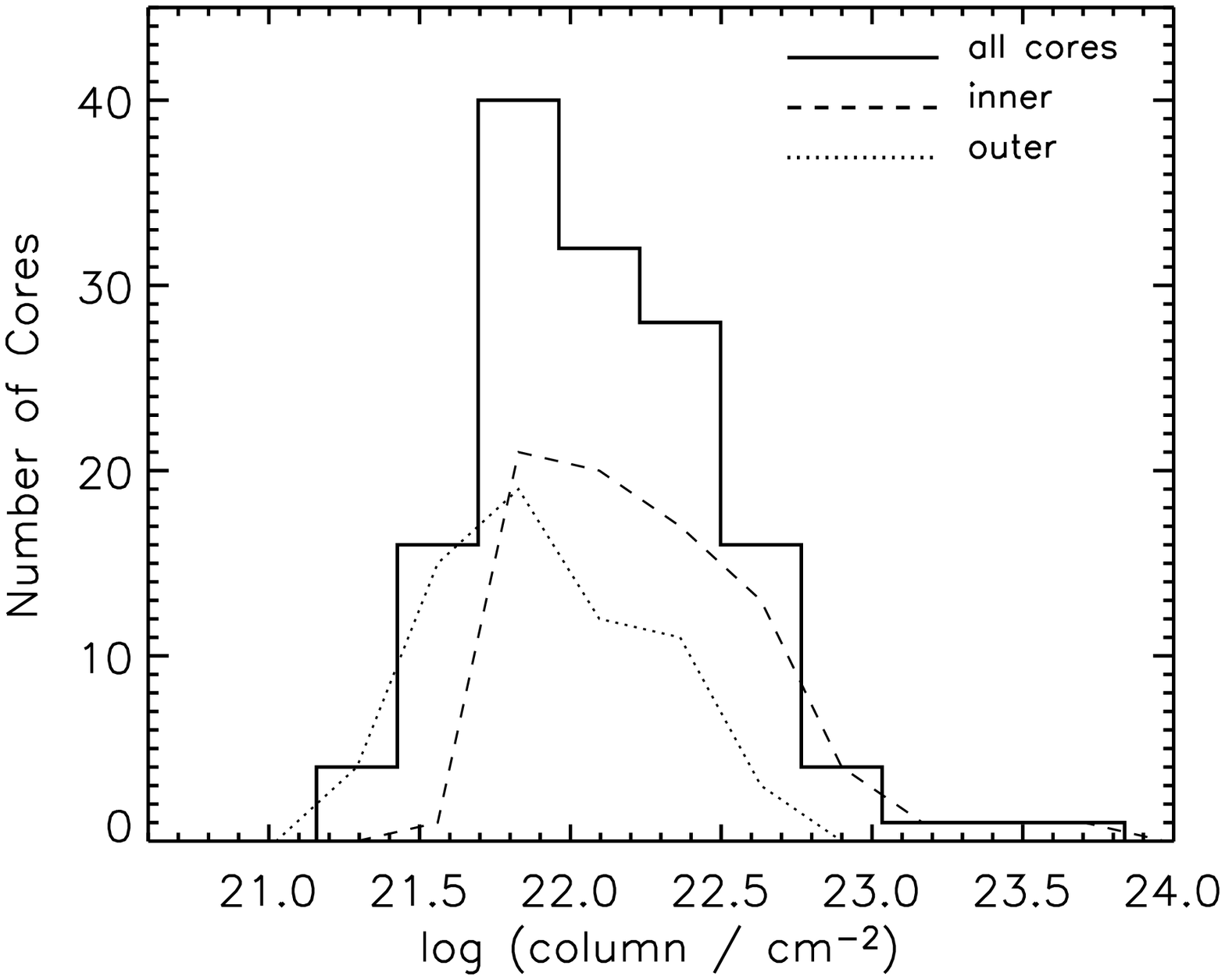,width=8.7cm}}
\centerline{\psfig{file=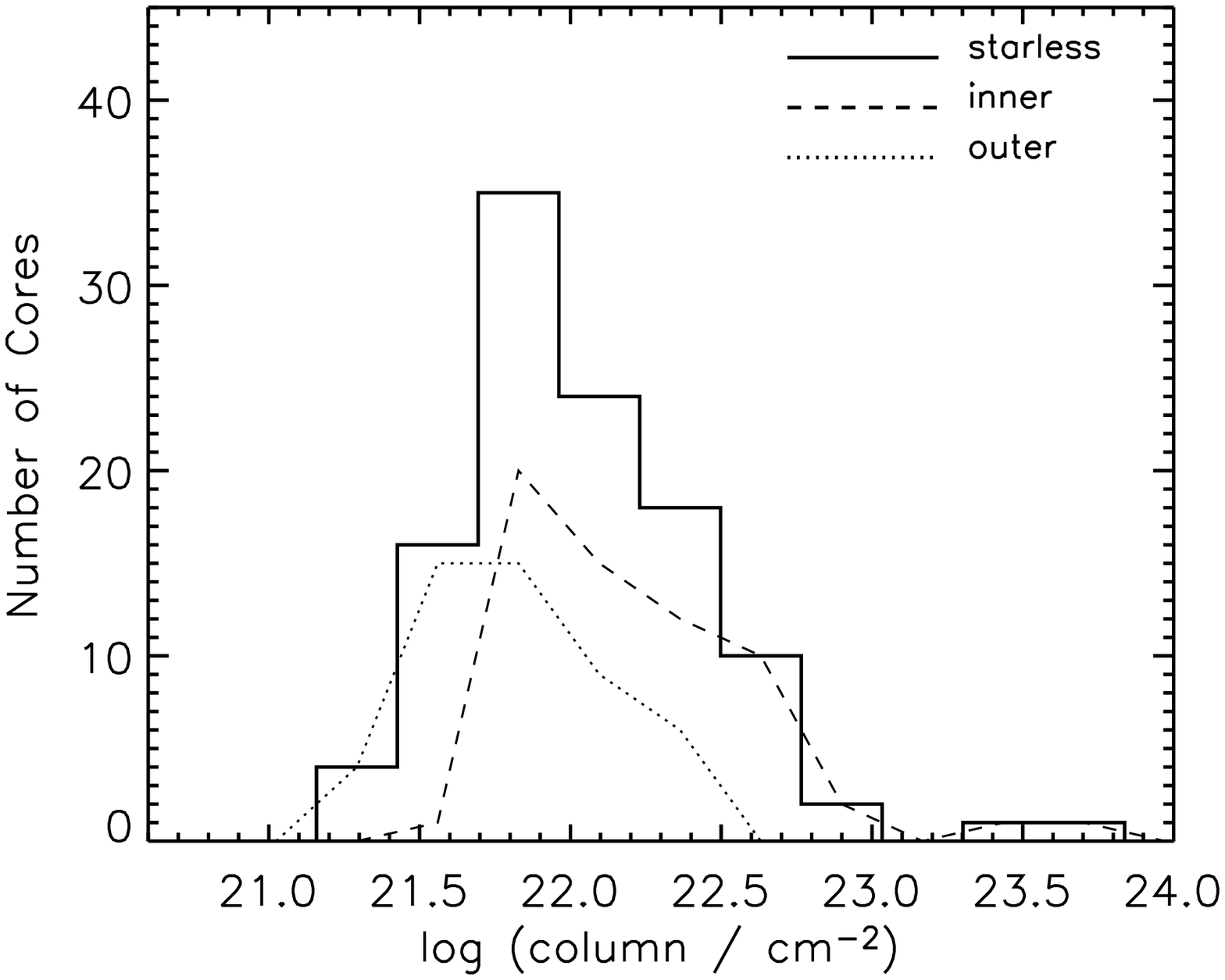,width=8.7cm}}
\caption{The number distribution of the mean column density of the cores
(also proportional to the surface pressure predicted
if a sub-sample were critical Bonnor-Ebert spheres, as given from 
Eq.~\ref{rad-be}). {\bf Upper Panel}: The full sample of 143 cores, with
79 within the inner crowded core and 64 external to it;
{\bf lower Panel:} the same distribution for just the starless cores (111
in total, 62 within the inner crowded core and 49 external to it).}
\label{pressure}
\end{figure}

The mass of each core is estimated on assuming it to be proportional
to the measured 1.2\,mm flux. The radiation is taken to be emitted from
cool dust grains and the intervening medium is optically thin. The
standard formulae for black bodies then yield the core masses from the
fluxes $S_{1.2mm}$ expressed in Janskys (e.g MAN98 and J00):
\begin{multline}
\label{eqn_flux_mass}
M_c = 0.58 \times  S_{1.2mm}~~ 
     \left[\frac{\kappa_{1.2mm}}{0.005~{\rm cm}^2~{\rm g}^{-1}}\right]^{-1} \cdot \\
      \left[ \exp \left(\frac{12~{\rm K}}{T_d}\right) - 1 \right]
      \left[\frac {D}{130~{\rm pc}}\right]^2 {\rm M}_\odot.  
\end{multline}
Here, we take a fixed opacity at 1.2\,mm of $\kappa$ = 0.005~cm$^2$~g$^{-1}$
and a fixed dust temperature of $T_d$ = 20\,K. It should be noted that these
quantities may vary across the cloud by a factor of order two, as fully
discussed by MAN98 and J00.

The derived masses are plotted as a function of core size in
Fig.~\ref{massrad}.
Here, we do not attempt to deconvolve the apparent cloud 
dimensions and define the mean observed clump angular radius of 
$\theta$ = 0.5~(FWHM$_{maj}$~$\times$~FWHM$_{min}$)$^{1/2}$ from the
FWHMs of the major and minor axes given in Table~\ref{tab_sources}.

Although the inferred masses and radii are widely distributed,
a mass-radius correlation is evident in Fig.~\ref{massrad}.
However, it should be remarked that large, low-surface brightness features
might have been supressed by the observational procedure and data reduction,
as discussed above. Hence, the apparent
correlation is possibly only produced by a deficiency of extended, moderately
massive features. (It should be noted that the smallest core radii
presented in the diagram are at the limit of our resolution and subject to 
significant error). At small radii and masses, the lower envelope to the
distribution of sources reasonably well reproduces the size dependency of the
mass detection limit as argued above, ${\rm M}_{\rm lim} \propto {\rm R}$.
To investigate a possible mass-radius correlation, locii of constant
average column 
density, corresponding to $\log {\rm M} \propto 2 \log {\rm R}_c$, are
overplotted in Fig.~\ref{massrad}. This demonstrates that at low radii and
masses the detection limit  yields a lower envelope to the
distribution which influences the correlation.

Splitting the sample into cores within the inner, crowded region (marked by
crosses, corresponding to the region investigated by MAN98 and J00) and those
spread over the extended cloud (diamonds) hints at
distinct lower envelopes for cores in the inner and outer regions, 
corresponding to average atomic hydrogen columns
of $8 \times 10^{21}$\,cm$^{-2}$ (inner region) and 
$4 \times 10^{21}$\,cm$^{-2}$ (outer regions). 
We believe that this is at least partially due to observational
limitations, as low surface-brighntess features are hard to identify
in the crowded central region.

The number distribution for the core columns is displayed in
Figure~\ref{pressure}. This demonstrates that the offset in column density
between the inner and outer regions is real and extends to the entire
distribution and not just the lower bound to the column density.
It is not caused only by a (possibly obervationally introduced) deficiency
of low column density cores in the inner region, but also by the absence
of high column density cores in the outer region, which would have been
easily detected.
The inner cores possess a systematically higher column density by a factor of 
approximately two on average. 

This suggests that a form of segregation, present in stellar clusters, is
also taking place in the clumps in $\rho$ Ophiuchus. However, as we will
demonstrate in Section \ref{chap_massdist},
the mass distribution is very similar in the inner and outer regions. 
The segregation is not in the mass but in the column density or pressure, 
hence compactness
of the cores. That is, the cores must be somewhat more compact to survive
in the crowded regions. This could also suggest that cores, in  general,
cannot be treated as isolated objects although the compacter cores, which
exist sufficiently long, may have reached a relatively independent state.
This scenario, however, must be reconciled with the dynamical time scales 
\citep{2001fdtl.conf..313B} which allow little time for core-core interactions.

The existence of two exceptionally massive, high density cores in the inner
region can be considered to be the result of physical processes rather than
core overlap along the line of sight. As can be seen from Fig.~\ref{massrad},
these two cores are both massive and distinctly compact and so lie in a
distinct region of the mass-radius diagram. These are the cores MMS001 
(known as \object{SM1} and \object{SM1N}) and MMS004 (\object{SM2}) within
the most crowded Clump A.
They appear to outline  a cavity around the bright, young \object{star S1}
\citep{1973ApJ...184L..53G}. This suggests that they consist
of swept-up and compressed gas, locally triggered into the collapse 
process.

Note also that restricting the analysis to just the starless cores
(Figs.\ \ref{massrad} and \ref{pressure}, lower panels) does not
alter the above results. However, the majority of the very low mass
compact cores, both within and outside the crowded region, already contain
young stellar objects. This suggests that such low-pressure cores only form
or survive by being gravitationally bound to a point source.

\subsection{Core Support}
Are the cores ephemeral or in a prolonged state, close to equilibrium?
Some cores could be internally supported by a uniform thermal, 
turbulent  or magnetic pressure and contained by a roughly equal external
pressure. In such two-phase media, the ambient medium may be quite diffuse,
warm atomic gas.
However, this cannot apply to the majority of the cores since the
internal density generally rises significantly towards their centres (MAN98).
Nevertheless, an ambient medium of constant pressure could be maintaining
a few of the diffuser low-mass cores in the outer regions. With a limited
variation in core temperatures,
uniform pressure would roughly lead to a relationship of the form
$\log {\rm M} \propto 3 \log {\rm R}_c$, i.e., steeper than
that corresponding to cores of constant column. This is clearly not
observed. 

Alternatively, the derived masses can be compared to those expected from
isothermal spheres
of the same size which are in hydrostatic equilibrium. In particular, there is
a maximum cloud mass which is stable to perturbations in pressure 
\citep{1955ZA.....37..217E,1956MNRAS.116..351B}. The maximum
stable mass of a Bonnor-Ebert sphere, which assumes self-gravitating,
isothermal cores bounded by an external surface pressure,
is $M_{BE} = 1.18~\sigma^4/(G^{3/2}P_s^{1/2})$ where $P_s$ is the 
surface pressure and  $\sigma$ is the one-dimensional velocity dispersion,
here equivalent to the isothermal sound speed \citep{2002A&A...394..275G}.  
Evaluating in terms of a typical molecular cloud pressure, we obtain
\begin{equation}
M_{BE} = 1.2 \left[ \frac{T}{20\,{\rm K}} \right]^2
      \left[ \frac{P_s}{2~~10^{-10}~~ {\rm dyne\,cm}^{-2}} \right]^{-1/2}~
         {\rm M}_\odot.
\label{massbe}
\end{equation}
In terms of size, the equilibrium sphere will be stable if the radius exceeds
the critical radius $R_{BE} = 0.41~G~M_{BE}/\sigma^2$. 
Assuming a fixed core temperature (MAN), we thus predict a shallower
relationship $\log {\rm M} \propto \log {\rm R}_c$ if the cores are close to 
the critical state.

In a similar fashion, we could assume that the surface pressure is quite
uniform and allow the temperature of the cores to vary by a small factor,
consistent with Eq.~\ref{massbe}. Then, eliminating $\sigma$ we obtain the
critical radius
\begin{equation}
R_{BE} = 5.1 \times 10^3 \left[ \frac{M_{BE}}{1~~ {\rm M}_\odot}\right]^{1/2}
\left[ \frac{P_s}{2~~10^{-10}~~ {\rm dyne\,cm}^{-2}} \right]^{-1/4} ~{\rm AU},
\label{rad-be} 
\end{equation}
which predicts the relationship $\log {\rm M} \propto 2 \log {\rm R}_c$.

The data, displayed in  Fig.~\ref{massrad}, indicate a wide range in core
masses for any given core size. Under the strict assumption that all cores
are {\em critical} Bonner-Ebert spheres, have the same temperature and only
thermal internal motions, this would imply a range in surface pressures 
exceeding 100.  The loci of constant mean column
presented in Fig.~\ref{massrad} also correspond to surface pressures
of $2 \times 10^{-10}$ dyne~cm$^{-2}$ 
(upper left line), $2 \times 10^{-11}$ dyne~cm$^{-2}$ (middle line) and 
$2 \times 10^{-12}$ dyne~cm$^{-2}$ (lower right line).
In contrast, J00 found a much narrower mass-radius correlation for cores in the
inner region, consistent with a smaller range in surface pressure.

\begin{figure}[th]
\centerline{\psfig{file=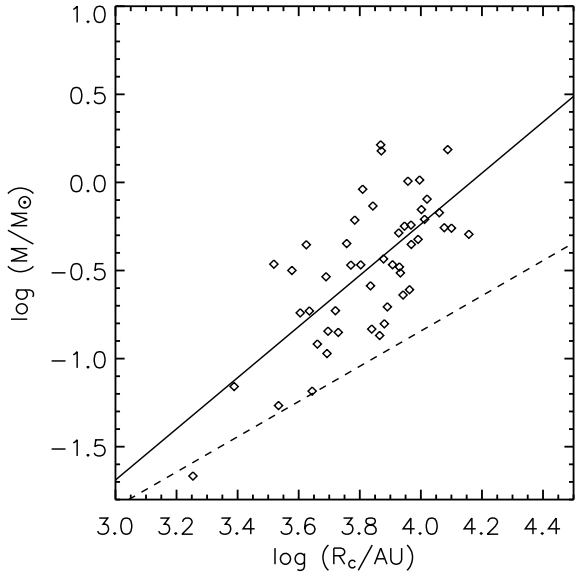,width=8.7cm}}
\caption{The mass-radius relation for the starless cores in the outer
regions. The solid line is the least squares fit and the dashed line is
suspected to be the detection limit with  ${\rm M} \propto {\rm R}_c$.  }
\label{mrstarless}
\end{figure}

The core mass is significantly less sensitive to the core size than that
predicted by a constant column. Least squares fits of the form
\begin{equation}  
\log ({\rm M}/{\rm M}_\odot) =  a~\log ({\rm R_c}/10,000~{\rm AU}) + b
\end{equation}
yield $a = 1.11 \pm 0.18$ and $b = -0.16 \pm 0.06$ for all starless cores
(correlation coefficient is 0.50) with $a = 1.45 \pm 0.22$ and
$b = -0.24 \pm 0.05$ for the outer starless cores (correlation coefficient
is 0.70) and  $a = 1.49 \pm 0.30$ and $b = 0.07 \pm 0.12$ for the inner
starless cores (correlation coefficient is 0.51).
Figure~\ref{mrstarless} displays the mass-radius relation for just the outer
starless cores. Obviously, the correlation is not very stringent, as also
indicated by the fairly low correlation coefficients. The offset between the
power laws for the inner and outer cores corresponds to a difference in size by
a factor of $\sim$1.2--1.3, which using Equn.\ \ref{rad-be} translates to a
pressure difference of a factor of $\sim$2--3. This is significantly less than
the surface pressure variations which can be inferred from the range of core
masses at some given radius or as have been derived by J00. Random pressure
variations within the inner and outer cloud gas seem to dominate the systematic
increase in pressure when going from the outer to the inner part, as one could
expect for a medium being subject to random, turbulent motions. 


Although the observational limit tends to flatten the mass-radius relation,
we cannot exclude an underlying intrinsic law of the form
$\log {\rm M} \propto \log {\rm R}_c$, as given by critical Bonnor-Ebert
spheres with a constant velocity dispersion. In this case, however, the wide
range in masses implies a range of 100 in the surface pressures to maintain
the cores in equilibrium, as given by  equation~\ref{massbe}.
This variation in pressure could be present, resulting from highly
supersonic turbulent motions (see \S~\ref{chap_concl}).
In fact more likely would be a steeper relation than indicated by the fit,
approaching $\log {\rm M} \propto 2 \log {\rm R}_c$, as expected for
critical Bonnor-Ebert spheres immersed in a constant external pressure
medium having varying velocity dispersions; due to the $\sigma^4$ dependence
of ${\rm M_{\rm BE}}$ only moderate variations would be needed to create the
more than 1 order of magnitude dispersion of mass at a given radius.
Again, this would more likely be a sign for turbulent in addition
to thermal motions, as it appears unlikely that starless cores exhibit large
enough temperature variations. Moreover, in a turbulent medium, cores will be 
continually forming and dispersing, and may 
temporarily resemble bound cores. In recent gravo-turbulent simulations, starless 
cores  were generally gravitationally unbound, suggesting that gravitational collapse 
occurs promptly after gravity becomes dominant \citep{2005ApJ...620..786K}.

\section{Spatial Distribution 
\label{chap_distrib}}         

\begin{figure}[th]
\centerline{\psfig{file=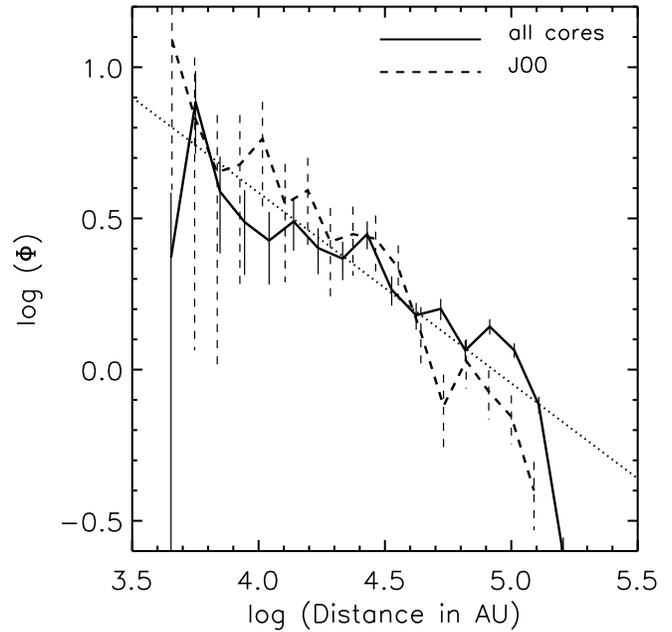,width=9.6cm}}
\caption{The two point correlation function as defined in the text
  for the $\rho$~Ophiuchus cores. The error bars correspond to $\surd$N
  statistics. The dotted straight line corresponds to a power-law of the form 
  $\omega \propto {\Delta}r^{-0.63}$.
  }
\label{correlate}
\end{figure}

The identified cores are strongly clustered at first sight. To quantify this,
we employ a two-point correlation function treating each core as a point
object in space. Following J00, we determine the number of core pairs,
$H_d(r)$, with separation between $\log (r)$ and  $\log (r) + d \log (r)$.
This is compared to the predicted distribution for a random sample of
cores spread over the apparent volume, $H_r(r)$. The two-point
correlation function is then defined as
\begin{equation}
\Phi = \frac{H_d(r)}{H_r(r)}  -   1.
\end{equation}
For the random sample, we consider a random surface distribution across a
sheet of the survey size and shape. Fig.\,\ref{correlate}
shows that the correlation function is positive on scales under $\sim 10^5$~AU.
For comparison, the data obtained by J00 are also plotted on
the Figure.

The survey size is seen to influence the correlation with the present
survey displaying more large scale structure while the J00 survey detected more
smaller cores, as expected from the observational constraints. 
J00 found that a power law of the form
$\Phi \propto {\Delta}r^{-0.75}$ fitted the data well out to 
$3 \times 10^4$~AU. The new survey suggests that this relationship can be
extended to above $10^5$~AU. Therefore, in contrast to J00, we exclude the existence
of a preferential scale near to $3 \times 10^4$~AU, which could correspond to
a Jeans length. Instead, the most obvious possible cause of hierarchical clustering
is turbulence. Supersonic turbulence would also generate a
wide range in pressure and column density, as required to interpret the
mass-size distribution (see \S~\ref{chap_concl}).

The inferred power law index of -0.63 is somewhat flatter than found by J00,
although within the error bars except on the largest comparable scales. Note
that, as remarked by J00, this value is close to the value measured for
galactic clustering of -0.668 for scales under 1 $^\circ$
\citep{1990MNRAS.242P..43M}. Furthermore, the function $\Phi$ is closely
related to the mean surface density of companions (MSDC), being approximately
proportional for values of $\Phi$ exceeding unity \citep{1997ApJ...482L..81S}.
Thus, it is interesting that the power law radial  dependence of the MSDC for
pre-main-sequence stars in \object{Taurus-Aurigae} is $\sim -0.6$ for scales 
$r > 0.04$\,pc (and corresponds to a fractal point distribution with 
index 1.4). The lower limit separates the bound systems from the unbound
stars, and was suggested by \citet{1999ApJ...525..440L} to correspond to both
a Jeans length and the size of bound molecular cores. 

For samples of Ophiuchi stars, \citet{1997ApJ...482L..81S} found an MSDC index 
of -0.5\,$\pm$\,0.2 above a break at 5000\,AU. On the other hand,
\citet{1998ApJ...497..721N} found  -0.36\,$\pm$\,0.06. As noted by
\citet{1998MNRAS.297.1163B}, flattening of the
correlation is expected on a timescale of 10$^5$\,yr, as unbound stars separate
and the system expands. 

\begin{figure}[th]
\centerline{\psfig{file=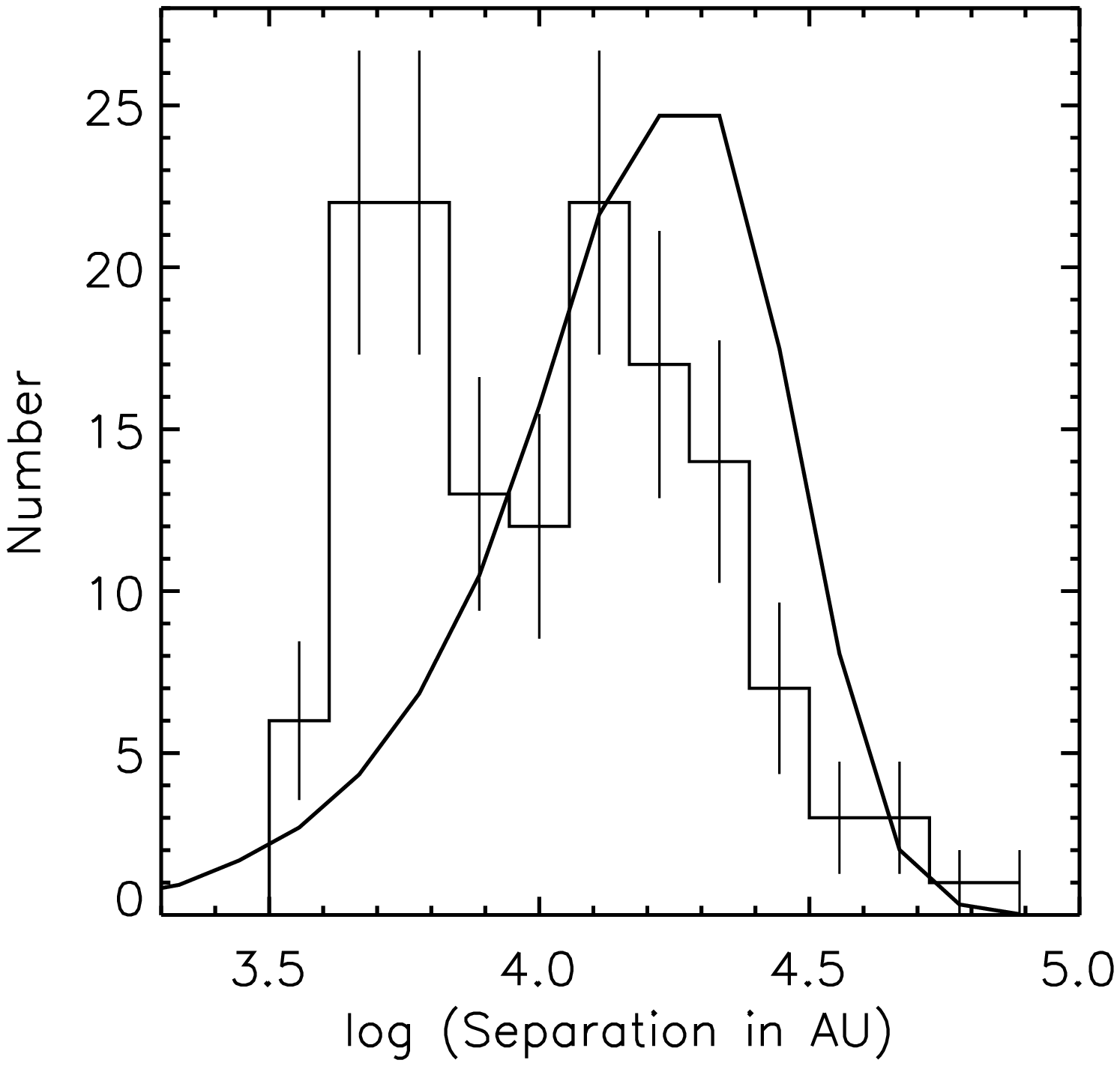,width=8.7cm}}
\centerline{\psfig{file=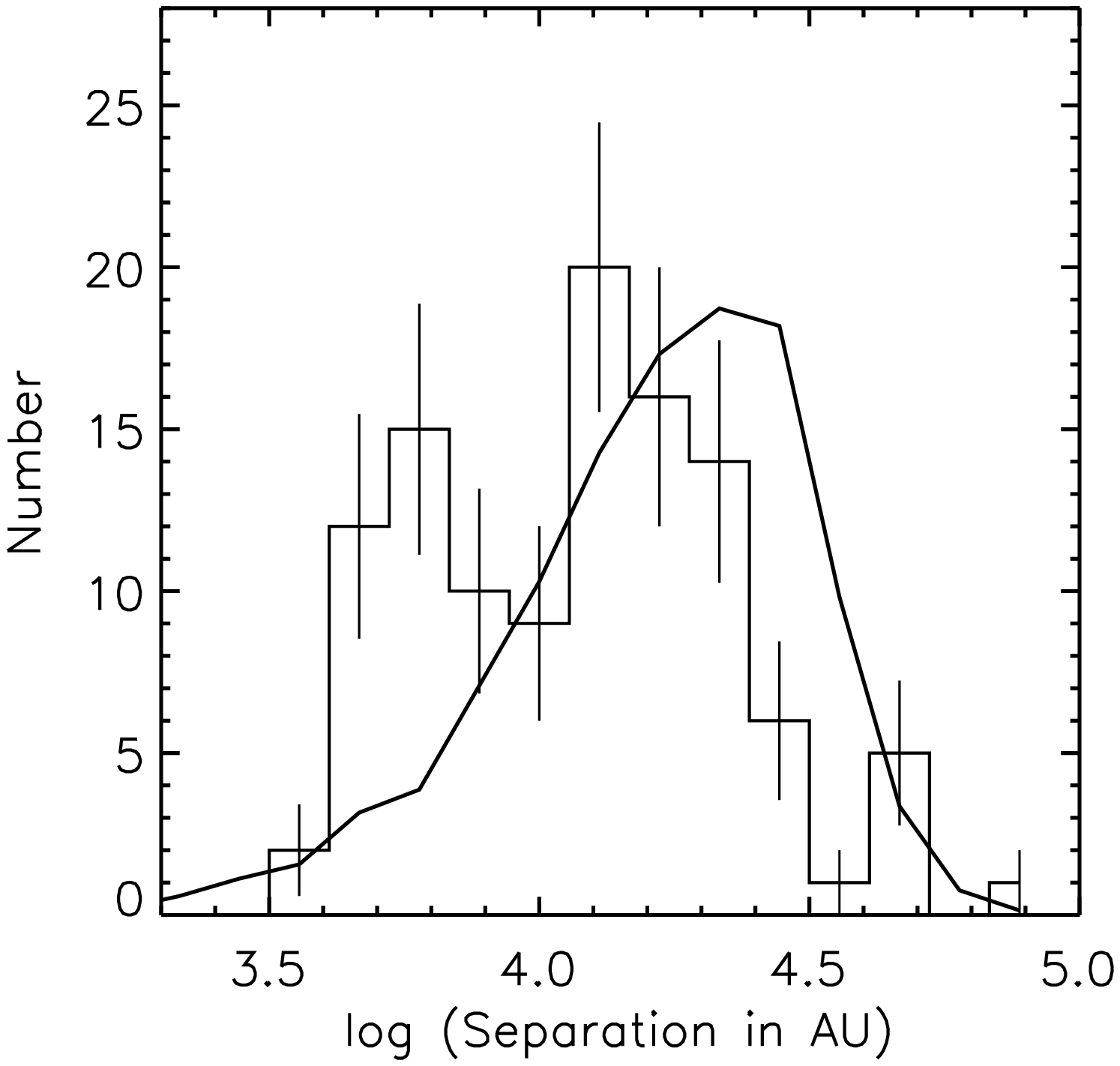,width=8.7cm}}
\caption{The number distribution of separations between neighbouring cores for
  all cores (upper panel) and the starless cores (lower panel). The error bars 
  correspond to $\surd$N statistics. The superimposed lines correspond to the
  predicted mean numbers for a random sample. The mean separation values are
  15,400~AU (all cores) and 17,200~AU (starless cores).}
\label{neighboursep}
\end{figure}

The frequency of separation between {\em neighbouring cores} complements the
two-point correlation function since it contains information on all orders
of the correlation functions.
The mean neighbour separation might also be interpreted as a typical Jeans
fragmentation length \citep{1993AJ....105.1927G,1995MNRAS.272..213L}. A length
scale could be apparent which might be hidden in the distribution involving
all pairs. In fact, the mean separation between neighbours in our starless
core sample is 17,400~AU and Fig.~\ref{neighboursep} indicates that there is
indeed a frequency peak near this separation distance. For comparison, a
random sample would have a mean separation of 22,100~AU (we simulated 500 sets
of randomly distributed starless cores spread through the entire region).
The random sample also indicates that we cannot interpret the peak in  the
separation frequency as being derived from som physical process which 
generates a break (knee) in a power-law functions with a steep decline
at large distances since a turnover is expected even in random data.

Of much more significance is the second frequency peak at low separations
(5000~AU) prominent in both panels of Fig.~\ref{neighboursep}, appearing
as if the hierarchical core clustering is modified by the fragmentation of
some close cores. In fact, this separation distance corresponds well to the
size of the cores themselves (see Fig.~\ref{massrad}). Hence, it is not
surprising that spatial structure is present on this scale.

\section{Mass Distribution  
\label{chap_massdist}}      

\begin{figure}[th]
\centerline{\psfig{file=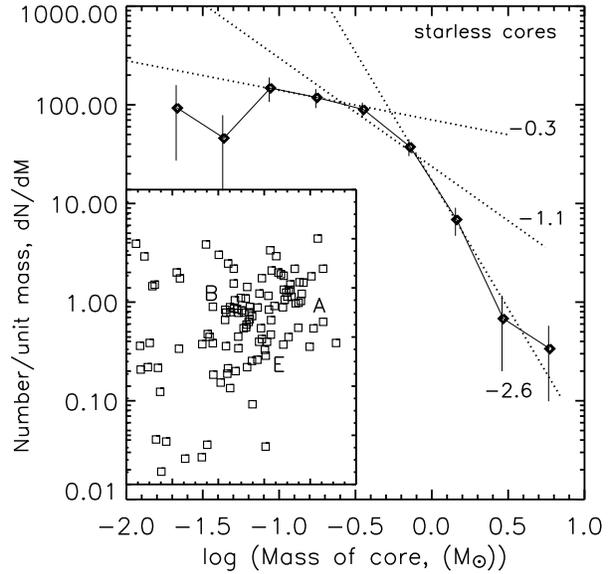,width=8.7cm}}
\caption{The number distribution by mass of the 111 starless cores in 
   $\rho$ Ophiuchus. The inset displays their locations. The dotted lines are
   the power law fits of the form dN/dM $\propto$ M$^{\alpha}$.
   Breaks in the power laws occur around
   0.1--0.3~M$_\odot$ and around 0.5--1~M$_\odot$. The Salpeter mass function
   would have a slope of $\alpha = -2.3$ in this format. The error bars
   correspond to $\sqrt{N}$ counting statistics. The inset displays the core
   positions in a wide region stretching 1$^\circ$ $\times$ 1.2$^\circ$, with
   the locations of the clumps A, B and E marked. }
\label{mf-starless}
\end{figure}

\begin{figure}[th]
\centerline{\psfig{file=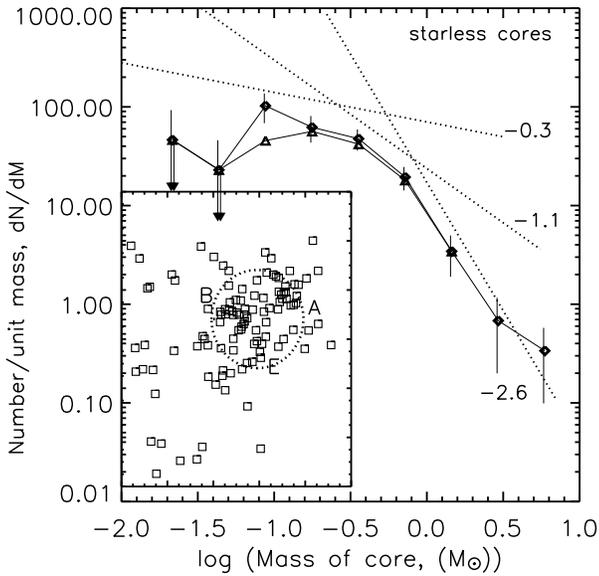,width=8.7cm}}
\caption{The number distribution by mass and location of the cores in 
   $\rho$ Ophiuchus for 111 starless cores. The dotted circle drawn in
   the inset divides the cores into an inner zone of 62 objects (diamond 
   symbols) and an outer area of 49 objects (triangles). The dotted lines,
   error bars and inset are as described in the Fig.~\ref{mf-starless} caption. }
\label{mf-circle}
\end{figure}

The distribution of core masses is displayed in Fig.~\ref{mf-starless} for all
the starless cores in the sample.
We also display in Fig.~\ref{mf-circle}, the mass functions for the inner (62)
and outer (49) starless samples, defined by a circle of radius 0.2$^\circ$,
located so as to encompass the crowded inner region studied by MAN98 and J00.
We find that a similar  mass function to that found by MAN98 and J00 for
the inner zone applies to the entire region. 

The core mass function appears to be hard to fit with a single
power law, being steeper at higher masses and flattening continously
towards lower masses. Power law fits with one or two breaks are
reqired for a satisfying fit.
The two-break power law fit as displayed in Figs.\ \ref{mf-starless}
and \ref{mf-circle} with two breaks is comparable to the 
approximation for the  stellar initial mass function (IMF) derived by 
\citet{2001MNRAS.322..231K}, where the power law indices are -0.3, $-$1.3,
and $-$2.3, although with considerable potential variation. Note that the
flatter power law holds for core masses below $\sim$ 0.14~M$_\odot$, in
comparison to the value of 0.08~M$_\odot$ quoted for the stellar IMF
\citep{2001MNRAS.322..231K}.

If we suppose that the break at low core masses is real and indeed occurs at
around 0.14~M$_\odot$, then the shift to a 0.08~M$_\odot$ break for the
stellar mass function has to be explained. On the one hand, it could be due
to errors in the assumptions which go into the core mass determination.
Although a closer distance would lower the derived core masses,
we have already assumed a quite close distance to $\rho$ Ophiuchus.  A larger
opacity would also reduce the core masses: an opacity of 0.009~cm$^2$~g$^{-1}$
would move the first break point to 0.08~M$_\odot$ for the distance of 130~pc.
However, a somewhat lower dust temperature than the assumed 20~K in the
flux-to-mass conversion would tend to shift the break point to higher masses.
Alternatively, if we assume the first break and the opacity are accurate, we
can speculate that just about one half of a core mass ends up constructing the 
star. The other half is dispersed in (i) jets (up to 30\%), (ii) dispersal 
by jet impact, (iii) early stellar and disc winds and (iv) other stars, brown
dwarves and planets.

\citet{2001A&A...372..173B} infer a 2 component power law IMF for Class~II
young stellar objects in $\rho$ Oph from ISOCAM data. The break 
the IMF is seen at around 0.55~M$_\odot$ and separates power law slopes
with indices $-$1.35$\pm$\,0.25 and $-$2.7 below and above the break,
respectively. This
IMF is `statistically indistinguishable' from the core mass function derived
by MAN98 and is also  consistent with the present data sets.

Broken power law fits to the core mass function are not very well constrained,
concerning the number of breaks as well as their location. A more continously
changing distribution such as a log-normal function might do a better job.
In fact, \cite{ballesterosparedesetal2005} derive core mass distributions
from numerical models of turbulent molecular clouds, finding mass functions
similar to log-normals. As their technique of clump extraction is similar to
our approach (taking all density enhancements, whether bound or not) both
studies should be well comparable. Finding a core mass function similar to
the ones derived from simulations further supports the idea of turbulence
as the main agent in shaping the cloud.

Whatever the exact, detailed  shape of the core mass function is,
its general, overall behaviour resembles the stellar mass function.
The MAN98 and J00 surveys both concentrate on cores which are argued to be
likely gravitationally bound. Our survey includes more extended,
lower surface brightness features, which are less likely to be bound.
However, numerical simulations showed that the structure of transient
density peaks forming and dispersing in a turbulent cloud may closely
resemble hydrostatic, gravitationally bound objects 
\citep{2003ApJ...592..188B}, so even the cores identified by J00 and MAN98
might turn out not to be bound.
Any scenario which closely correlates the core and stellar masses
makes presumptions concerning the stratification of the envelopes and their
subsequent evolution (i.e. whether certain cores might collapse while others
disperse). To select cores according to such criteria requires a consistent
means of interpreting the data from cores of very contrasting sizes and masses.
Moreover, as pointed out by J00, such surveys are inevitably incomplete despite
our unbiased strategy. 
Given all these caveats, the resemblance of core and stellar mass functions
is even more remarkable. 

Possibly the most important result of our study is that
no significant difference is measured in the core mass function
between the inner and the outer regions where the functions overlap. 
As shown in Fig.~\ref{mf-circle}, the few highest mass cores lie in the 
inner region. 
Thus, while the compactness of a core varies in space,
the mass distribution does not. This must be reconciled with the fact that the 
inferred surface pressures are, on average, lower in the external region
which would imply a higher Jeans mass (neglecting
possible temperature differences).

\section{MMS126: a low-mass Class~0 candidate and its CO outflow 
\label{chap_mms126}}              

\begin{figure}[th]
\centerline{\psfig{file=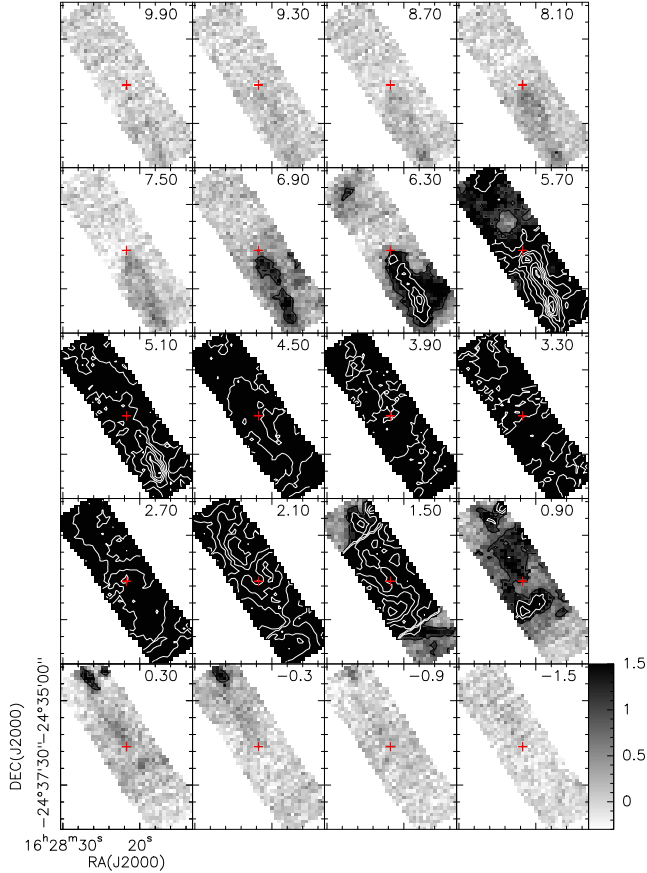,width=8.5cm,angle=270}}
\caption{CO J=3-2 velocity channel maps of the MMS126 outflow. The greyscale
is tailored to emphasize the low surface brightness outflow lobes. The contours
are spaced by 1~K.}
\label{fig_MMS126molflow}
\end{figure}

\begin{table*}
  \caption{Physical properties of the MMS126 molecular outflow. The values
    given in parentheses have been corrected for optical depth (assuming
    $\tau_{\rm CO}/(1-e^{-\tau_{\rm CO}}) = 3.5$) and inclination (assuming
    $i = 57.3^\circ$).
    }
  \label{tab_outflow}
  \begin{tabular}{lcccccc}
    \hline
       &  \multicolumn{2}{c}{Blue} & \multicolumn{2}{c}{Red} & \multicolumn{2}{c}{Total}\\
    \hline
    Maximum velocity (km/s)& 6 & (11.1) & 7 & (13) &&\\
    Size (AU/pc)& 20500/0.10 & (24400)/(0.12) & 19500/0.095 & (23200)/(0.11) & &\\
    Mass ($\times 10^{-4}$M$_\odot$) & 1.6 & (5.5) & 8.1 & (28.2) & 9.6 & (33.7)\\
    Dynamical timescale ($\times 10^3$~years)& 16.2 & (10.4) & 15.4 & (9.9) &&\\
    Momentum ($\times 10^3$~M$_\odot$ km/s) & 5.4 & (35) & 25.6 & (166) & 31.0 & (201) \\
    Momentum flux($\times 10^{-8}$~M$_\odot$ km/s y$^{-1}$) & 3.3 & (34) & 15.8 & (160) & 19.1 & (193) \\
    Kinetic Energy ($\times 10^{-3}$~M$_\odot$ (km/s)$^2$)& 0.98 & (12) & 4.6 & (55) & 5.6 & (67) \\
    Mechanical luminosity ($\times10^{-5}$L$_\odot$)& 1.0 & (19) & 4.9 & (92) & 5.9 & (110) \\
    \hline
  \end{tabular}
\end{table*}

MMS126 is a compact but resolved source forming part of a roughly east-west
oriented, filamentary structure. It appears to coincide
with a faint, cold IRAS source (\object{IRAS~16253$-$2429}: F12: $<$0.25~Jy; F25: 
$<$0.42~Jy; F60: 2.91~Jy; F100: 16.34~Jy). Close inspection of the HIRES
processed IRAS maps reveals a faint source at 25$\mu$m, which however
coincides with the nearby MMS060 = \object{WSB~60}. This latter source
is also detected by ISO at 7 and 14~$\mu$m \citep{2001A&A...372..173B}
but no clear source is seen at the position of MMS126. Finally, MMS126
appears to be driving a collimated near-IR H$_2$ outflow 
\citep{2004A&A...426..171K} and a molecular (CO) outflow 
(see Sect.\ \ref{chap_molflow} below).

Taking these arguments together (apparently cold FIR to millimetre source,
no clear counterpart at wavelengths shorter than 25$\mu$m, and association
with an outflow) we conclude that MMS126 is a very young stellar object,
possibly still in the Class~0 phase. MMS126 would then be the second
Class~0 object found in the $\rho$-Oph main cloud besides the
prototypical source \object{VLA1623} \citep{1993ApJ...406..122A} but is
of significantly smaller mass: using Eqn.\ \ref{eqn_flux_mass}, we
derive a circumstellar dust mass of about 0.2~M$_\odot$ 
(VLA1623: 0.6~M$_\odot$, this work; \citet{1993ApJ...406..122A}).

\subsection{The CO outflow        
\label{chap_molflow}}             
We observed the area around MMS126 in the CO(3-2) line using the 
JCMT\footnote{The James Clerk Maxwell Telescope is operated by The Joint
Astronomy Centre on behalf of the Particle Physics and Astronomy Research
Council of the United Kingdom, the Netherlands Organisation for Scientific
Research, and the National Research Council of Canada.} on
June 17 \& 30 and on July 30, 2005. We took two position switched on-the-fly
maps centered on MMS126, each covering a 80\arcsec{}$\times$180\arcsec{} field
at a position angle of 32$^\circ$, sampling every 5\arcsec{} with a 5\arcsec{}
spacing between rows. Smaller maps were added towards the end of the outflow
lobes. The
integration time on each point was 4~sec. The weather was good, with
a stable atmosphere and $\tau_{230~{\rm GHz}} \sim 0.055$ (June 17), 
$\sim 0.08$ (June 30), and $\sim 0.045$ (July 30).
In addition, spectra in frequency-switched mode were taken towards the
position-switch OFF position.
The spectra were baseline-subtracted, resampled to 0.6~km/s resolution,
the OFF-position spectrum added, and finally gridded into one datacube
within CLASS.

Figure \ref{fig_MMS126molflow} shows the resulting channel maps. Along
the outflow axis indicated by our near-infrared H$_2$ observations there
is clear evidence for a CO flow. Red-shifted emission is seen towards the
south-west and some blue-shifted emission towards the north-east.
Overall, the range over which higher-velocity gas is seen is fairly narrow.
In the south-western, red-shifted lobe, it extends over a few velocity bins
from about 5.7 to 8.5~km/s. Reasonably bright emission in the north-eastern,
blue-shifted lobe is only seen in its tip, otherwise it is hardly
separated and only faintly seen in two channels around 0~km/s; the contours in
the velocity bins around 1~km/s indicate that this lobe contributes
significantly at these velocities too, but is diluted by ambient emission.
Outflow emission extends over about 2\farcm5 (20000~AU/0.1~pc) out to the
edges of the map, with both lobes showing a somewhat brighter, compact CO blob
there, which might well be a terminating working surface; H$_2$ emission is
also seen out to a similar distance in the south-western lobe.

We have derived estimates of basic flow properties as follows:
the measured brightness temperature was converted to molecular gas
column densities assuming the CO to be optically thin and in LTE at a
temperature of 30~K; the CO abundance was assumed to be $10^{-4}$ of the
H$_2$ abundance. Masses were derived as a function of velocity for each 
0.6~km/s wide channel. Kinetic energies and momenta were derived as
$E_{\rm kin} = 1/2 \sum M(v)\cdot (v - v_{\rm cen})^2$ and
$P = \sum M(v)\cdot (v - v_{\rm cen})$, with $v_{\rm cen} = 3.4$~km/s.
A characteristic timescale was derived by dividing the maximum distance
at which high-velocity emission was seen by the maximum CO velocity
observed. The south-western, red-shifted lobe is seen to extend to the edge
of the mapped area, so a maximum distance of 90\arcsec{} was used; the 
north-eastern, blueshifted lobe seems to terminate within the mapped area.
Finally, mechanical luminosities and momentum input rates were determined
by dividing the total kinetic energy and momentum by the characteristic
timescale. 

Table \ref{tab_outflow} gives a list of outflow properties, for the
blue- and red-shifted lobes alone and for the entire outflow. In
parentheses we also list the outflow parameters after applying
corrections for optical depth effects and inclination, following
\citet{1996A&A...311..858B} (assuming
$\tau_{\rm CO}/(1-e^{-\tau_{\rm CO}}) = 3.5$ and an inclination 
$i = 57.3^\circ$).
The red- and blue-shifted emission does not seem to overlap, indicating
that the flow axis is neither very close to the plane of the sky nor
very nearly perpendicular to it, so $i = 57.3^\circ$ might in fact be a 
reasonable assumption.

A comparison with the outflow parameters listed in \citet{1996A&A...311..858B}
shows that the MMS126 molecular outflow is one of the less energetic flows,
as could be anticipated by the poor contrast of the flow against the ambient
cloud material even in the CO(3-2) line and its fairly narrow extent in
velocity space.
Its momentum flux rate is much more typical for a more evolved object than
for a Class~0 protostar. However, our map shows the flow to be fairly well
collimated, which is more typical for the very youngest molecular outflows.
Along with the deeply embedded, low  circumstellar mass, 
low-luminosity nature of the driving source, the more likely explanation
for the low energy/momentum input rate to the flow seems to be a very low 
protostellar mass (and age) rather than MMS126 being a relatively mature
object at the end of the accretion and outflow phase.

\section{Summary \& Conclusions        
\label{chap_concl}}         

We have performed a 1.2 millimetre survey of the main body of the 
\object{$\rho$ Ophiuchi star-forming cloud}. Our one square degree field enables us
to verify and extend the results from previous  surveys (MAN98 and J00).
We detect and measure 143 cores, including 111 with no detected stellar
objects. We identify a core, \object{MMS126}, which appears to harbour a new low-mass
Class~0 protostar, driving a CO molecular outflow.

By comparing core masses with their sizes (as measured by the flux and FWHMs),
we find that the cores contained within an inner circle of 0.2$^\circ$
are compacter. This implies they are generally denser and of higher pressure.
However, on comparison to the surface pressure required for hydrostatic
equilibrium, there appear to be vast pressure variations (by a factor 10 to
100) within each location.

The data indicate that the mean core mass increases with the core radius. This
can be measured despite the expected lower bound to the core masses given by
the observational sensitivity (which would tend to distort the relation
towards the form $\log {\rm M} \propto \log {\rm R}_c$)
and despite the very wide range in possible core masses at each radius.

The clustering has also been investigated by deriving the two-point
correlation function. We confirm previous results that the cores are
highly clustered, similar to that of galaxy clusters (J00). However, from
the two-point correlation function, we find no evidence for a critical
length to identify with a Jeans length. However, the frequency of nearest
neighbours provides a more local measure and reveals an abundance of small
cores with low separation (5000~AU) -- of order of the 
mean core size. This suggests that larger cores may fragment on this scale.

Moreover, the relevant velocity dispersion (in place of the
sound speed) will also vary across the cloud, possibly producing velocity
variations positively correlated with separation. Therefore, although the
Jeans length remains a relevant length scale in turbulent fragmentation and
collapse theory, an extended power-law correlation function could still be
expected in a turbulent environment.

The cores are not distributed isotropically. There is a strongly
preferred direction, transverse to the direction of the Sco OB association.
However, the orientations of the major axes of the individual cores are
consistent with a random distribution (\S \ref{chap_orient},
online version only).

The mass function may be approximated by a broken power law. As previously
noted by MAN98 for the inner regions, this is similar to
that of the galactic field IMF for stars. The flattening at lower masses
occurs at a mass of 
$\sim$0.1 to 0.3~M$_\odot$, about twice as high as the
equivalent break mass found for stars. We suggest that this difference
is consistent with the core mass loss expected during the star formation
process although the absolute value of the break is sensitive to
assumed mass conversion parameters. An upper break at $\sim$0.5 -- 
1.0~M$_\odot$ is also found.
A smoothly changing function such as a log-normal
might be better suited to describe the core mass function derived from
our observations, similar to the mass functions derived from numerical models
of turbulence driven clouds by \citet{ballesterosparedesetal2005}.

The core mass function does not vary with position. That is, there is no
evidence for a general mass segregation although the few highest core masses
are found in the inner crowded zone.

How have the cores in L1688 developed?  Turbulent motions are inherent to the
entire clumps \citep{1990ApJ...365..269L} and (although subsonic) also
to the cores \citep{2001fdtl.conf..313B}.
We may thus expect supersonic turbulence to drive
high pressure and density variations on the scale of the cloud and clumps.
Some density peaks within the chaotic flow may involve sufficient mass to
become gravitationally bound; \citet{2003MNRAS.338..817E} argue that this
scenario is the most likely one of the three they consider to have produced
the millimetre continuum sources in Ophiuchus. 
The turbulence then decays within these proto-cores although it is still
sufficient to generate molecular line profiles and contribute to core
support in the observed cores \citep{2001fdtl.conf..313B}. As found in
numerical simulations, the structure of these dynamical cores may closely
resemble hydrostatic, gravitationally bound objects 
\citep{2003ApJ...592..188B}; they should also possess prominent infall
signatures as indeed are found in several starless cores
\citep{2001fdtl.conf..313B}. Furthermore, it is very likely that the detected
cores are biased towards the slowest evolving long-lived objects. That is,
the cores we actually observe are not typical of those which
form but are simply a particular subset of cores
which  evolve very  slowly, perhaps not even prone to collapse into stars. 


The present study is consistent with a supersonic
turbulence interpretation through the wide range in pressures, the hierarchical
clustering and the randomness in core orientations, and the shape of the
mass function. Turbulence, however, is generally supposed to be a means of 
dynamical support and rapid dispersal rather than quiescent confinement of
equilibrium core configurations. If the clouds are close to equilibrium,
the linear  mass-radius relation would correspond to a fixed velocity
dispersion. This indicates that a transition to coherence has already taken
place and is a sign that the H$_2$ velocity dispersion is transonic or
subsonic in the cores \citep{1998ApJ...504..223G}. In recent hydrodynamic
gravoturbulent simulations \citep{2005ApJ...620..786K}, the majority of cores
are actually found to be coherent and 80\% are subsonic or transonic. The
cores are produced behind shock waves, at stagnation points in converging
turbulent flows. However, starless cores are also found to be gravitationally
unbound in these simulations. This should in the future be tested by 
measurement of the virial masses of a large sample of the starless cores.

Further implications of the above results will be explored in a following work
in which we relate the millimetre properties to those of the infrared and
optical outflows from protostars and young stars. In combination with 
molecular spectroscopic studies of the cores, we will be able to constrain
models for star formation in $\rho$ Ophiuchus.

\begin{acknowledgements}
TS thanks the Alexander von Humboldt Gesellschaft for support through
a Feodor Lynen fellowship.
MDS thanks DCAL, Northern Ireland. Thanks are due to Doug Johnstone for 
providing the J00 and J04 $\rho$-Oph SCUBA maps.
This research has made use of the SIMBAD database, operated at CDS, 
Strasbourg, France. The Digitized Sky Survey was produced at the Space 
Telescope Science Institute under U.S. Government grant NAG W-2166. The images
of these surveys are based on photographic data obtained using the Oschin
Schmidt Telescope on Palomar Mountain and the UK Schmidt Telescope. The plates
were processed into the present compressed digital form with the permission of
these institutions.
This publication makes use of data products from the Two Micron All Sky Survey,
which is a joint project of the University of Massachusetts and the Infrared
Processing and Analysis Center/California Institute of Technology, funded by
the National Aeronautics and Space Administration and the National Science
Foundation. 
\end{acknowledgements}

\bibliography{1331}

\Online
\appendix

\section{Notes on individual objects 
\label{app_indiv}}           

\begin{figure*}[th]
  \vspace*{1cm}
  {\bf available as jpg: 0511093.1331fa01.jpg}\\
  {\bf full postscript version available at http://www.ifa.hawaii.edu//users/stanke/preprints.html}
  \vspace*{1cm}
  \caption{The location, size, and orientation of the 143 millimetre sources
         identified from our survey.}
  \label{fig_findingchart}
\end{figure*}

\begin{figure}[th]
  \includegraphics[angle=-90,width=\linewidth]{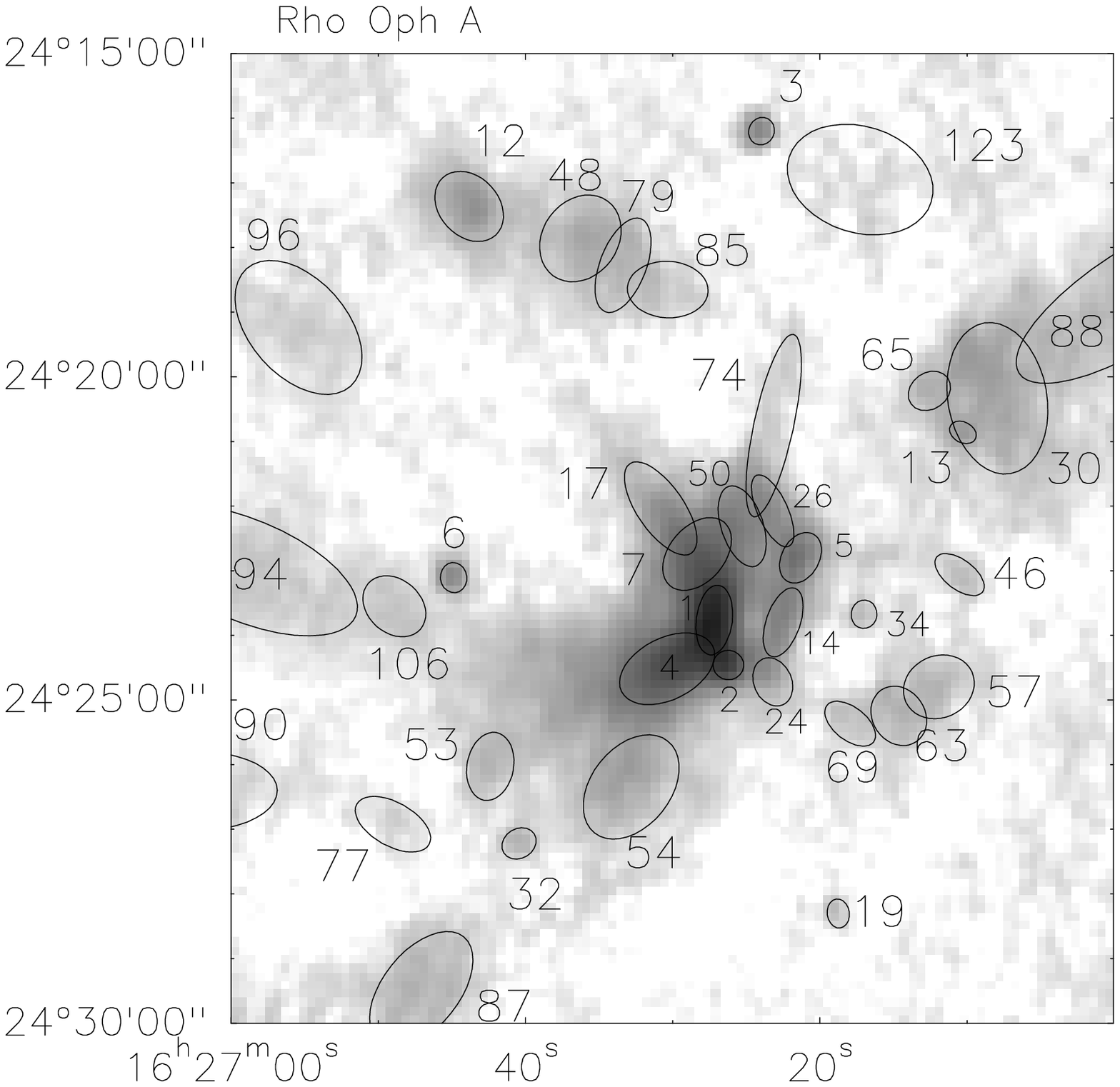}
  \caption{The location, size, and orientation of the millimetre sources
         in and around the molecular \object{clump A}.}
  \label{fig_findingchart_A}
\end{figure}

\begin{figure}[th]
  \includegraphics[angle=-90,width=\linewidth]{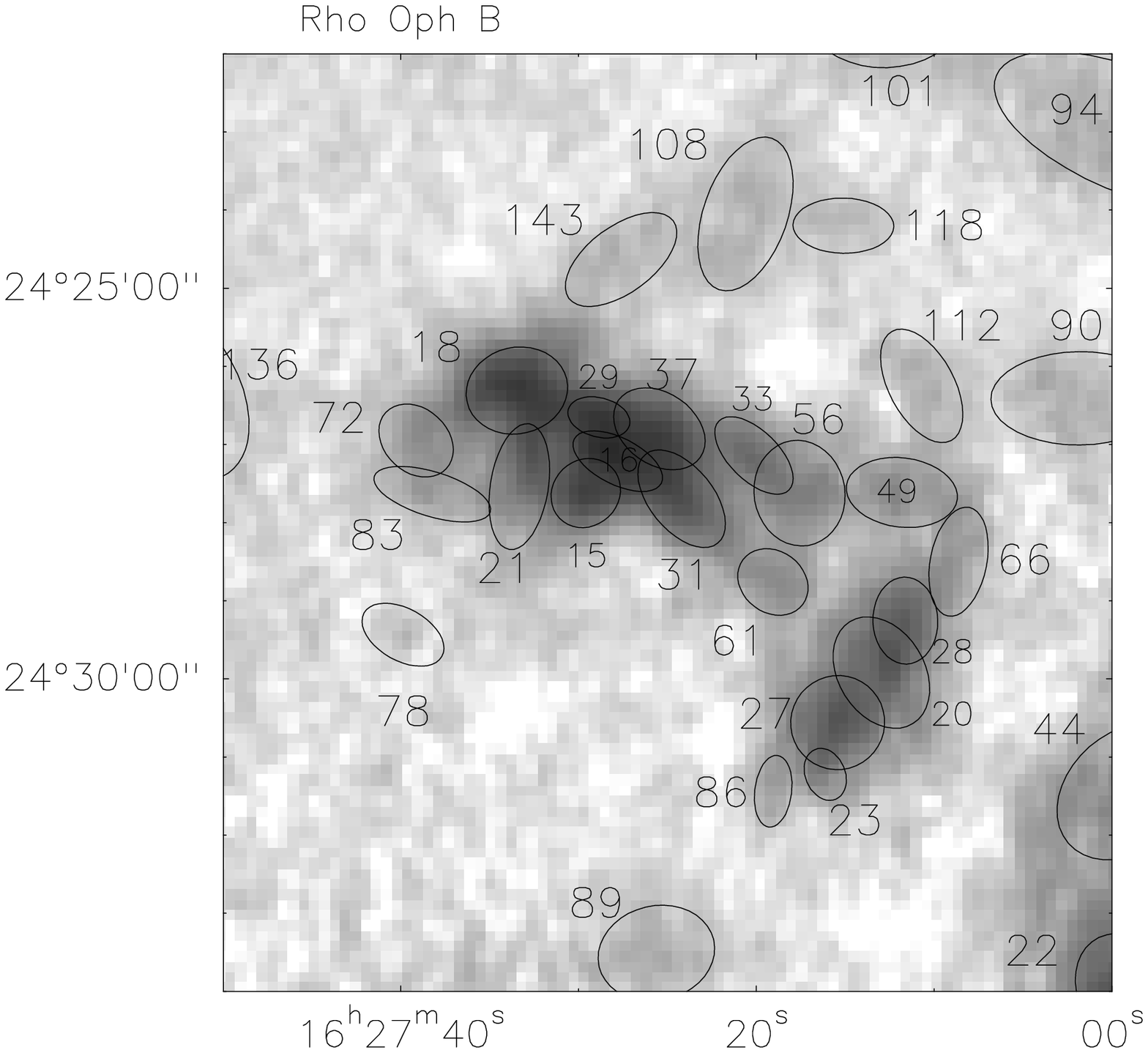}
  \caption{The location, size, and orientation of the millimetre sources
         in and around the molecular \object{clump B}.}
  \label{fig_findingchart_B}
\end{figure}

\begin{figure}[th]
  \vspace*{1cm}
  {\bf available as jpg: 0511093.1331fa04.jpg}\\
  {\bf full postscript version available at http://www.ifa.hawaii.edu//users/stanke/preprints.html}
  \vspace*{1cm}
  \caption{The location, size, and orientation of the millimetre sources
         in and around the molecular \object{clumps C, E, and F}.}
  \label{fig_findingchart_CEF}
\end{figure}

In this section we present a list of sources identified from our survey
(Tab.\ \ref{tab_sources}), along with finding charts (Figs.\ 
\ref{fig_findingchart}, \ref{fig_findingchart_A}, \ref{fig_findingchart_B},
and \ref{fig_findingchart_CEF}) and short notes on the properties of individual
sources, such as the presence of optical or near-infrared (stellar)
counterparts, or whether absorption features as seen on the DSS or 2MASS
images can be associated.

\begin{sidewaystable*}
\caption[]{List of 1.2~mm sources found in our survey. The sources are
listed in the order they were identified by the wavelet decomposition plus
{\em clumpfind} technique.
F$_{\rm int}$ gives the integrated source flux in mJy;
F$_{\rm peak}$ (a), (b), (c), and (d) give the peak fluxes
determined in the following ways:
(a): the maximum pixel value inside an ellipse defined
   by the minor and major axes of a gauss-fit to the source;
(b): as above, but using only the wavelet scales from where on
  the source is detected first (given in parenthesis);
(c): the peak flux after re-binning according to the source minor
axis (see Sect.\ \ref{chap_res} for details);
(d): the peak flux of the composite Gaussian model for the source. 
The columns labelled maj and min give the 
major and minor axis of a single Gaussian fit to the source (in arcseconds).
PA gives the position angle of the source major axis in degrees
east of north.   The columns labelled as
MAN98 and J00 give cross-identifications to the sources identified by 
\citet{1998A&A...336..150M} and \citet{2000ApJ...545..327J}, respectively. 
A long dash means that the source is outside the
respective survey area, and ``n.i.'' labels features which are visible in the
respective survey, but not identified as a separate feature.
Cross-identifications given in parentheses mean that a feature found here is
a part of a feature noted by MAN98 or J00. Finally, a comment on 
association with known YSOs is given. Apparently starless features are marked
as ``sl''.}
    \label{tab_sources}
\begin{tabular}{lccrrrrrrrrlp{2.1cm}l}
\hline
       &  RA        &   DEC     & F$_{\rm int}$ &\multicolumn{4}{c}{F$_{\rm peak}$ (mJy)} &\multicolumn{2}{c}{FWHM (\arcsec)}&PA& MAN98  & J00 &Comment\\
       &  J2000     &   J2000   &        (mJy)&(a)& (b) &(c) &  (d)  &  maj   &  min   &        &                 &            &  \\
\hline
MMS001 & 16:26:27.2 & -24:23:46 &  7775   &1999&1999 (1)&1828&  2002 &  65    &  33    &  -7    & \object{SM1}, \object{SM1N}       & \object{16264-2423} & sl (\object{SM1}, \object{SM1N})\\
MMS002 & 16:26:26.2 & -24:24:28 &  1205   & 924& 924 (1)& 864&   912 &  29    &  27    &  85    & \object{VLA1623}         & \object{16264-2424} & YSO: \object{VLA1623} Class 0\\
MMS003 & 16:26:24.0 & -24:16:12 &   242   & 248& 248 (1)& 227&   251 &  26    &  24    & -28    &  ---            & \object{16264-2416} & YSO: \object{YLW32} opt/IR TTS\\
MMS004 & 16:26:30.4 & -24:24:31 & 10255   & 973& 973 (1)& 912&   987 &  95    &  57    & -63    & \object{SM2}             &     n.i.   & sl\\
MMS005 & 16:26:21.3 & -24:22:48 &  1060   & 357& 357 (1)& 320&   351 &  50    &  35    & -29    & \object{LFAM1}           & \object{16263-2422} & YSO: \object{LFAM1}; Class I?\\
MMS006 & 16:26:44.9 & -24:23:06 &   291   & 230& 230 (1)& 217&   226 &  28    &  24    &   6    &  ---            & \object{16267-2423} & YSO: \object{GSS39} IR star Class I?\\
MMS007 & 16:26:28.3 & -24:22:45 &  3137   & 490& 490 (1)& 425&   482 &  77    &  51    & -41    & \object{A-MM6}/\object{7}        &    n.i.    & sl\\
MMS008 & 16:26:58.4 & -24:45:36 &   228   & 145& 145 (1)& 133&   153 &  30    &  24    &  59    &  ---            & \object{16269-2445} & YSO: \object{SR24} TTS binary\\
MMS009 & 16:27:09.1 & -24:37:25 &   316   & 154& 154 (1)& 129&   158 &  34    &  30    &  -1    & \object{Elia 2-29}       & \object{16271-2437b}& YSO:\object{Elia 2-29}\\
MMS010 & 16:26:23.7 & -24:43:12 &   153   & 132& 132 (1)& 119&   142 &  26    &  25    & -32    &  ---            &    ---     & YSO: \object{YLW34} TTS\\
MMS011 & 16:27:27.3 & -24:40:55 &   967   & 193& 193 (1)& 138&   193 &  82    &  58    &  79    & \object{IRS43}           & \object{16274-2440} & YSO: \object{WLY 2-43} Class I\\
MMS012 & 16:26:43.8 & -24:17:22 &  1109   & 205& 205 (1)& 170&   201 &  73    &  55    &  43    &  ---            & \object{16267-2417} & sl\\
MMS013 & 16:26:10.3 & -24:20:52 &    84   &  70&  70 (1)&  70&    78 &  26    &  19    &  62    & \object{GSS26}           & \object{16261-2420} & YSO: \object{GSS26} Class I?\\
MMS014 & 16:26:22.5 & -24:23:48 &   964   & 265& 265 (1)& 254&   261 &  67    &  32    & -19    & \object{A-MM1/2/3}       &    n.i.    & sl\\
MMS015 & 16:27:29.6 & -24:27:37 &  1336   & 284& 284 (1)& 246&   283 &  56    &  51    & -46    & \object{B2-MM10}         & \object{16274-2427c}& YSO: \object{VSSG17} (Class I?)\\
MMS016 & 16:27:27.8 & -24:27:13 &  1058   & 236& 236 (1)& 220&   240 &  75    &  36    &  63    & \object{B2-MM8}          &(\object{16274-2427b})& YSO: contains NIR YSO \object{VSSG18}\\
MMS017 & 16:26:30.8 & -24:22: 2 &  1145   & 187& 187 (1)& 163&   202 & 101    &  43    &  35    &  n.i.           &    n.i.    & sl\\
MMS018 & 16:27:33.5 & -24:26:19 &  2900   & 350& 350 (1)& 307&   333 &  78    &  66    & -77    & \object{B2-MM16/17}      & \object{16275-2426} & sl\\
MMS019 & 16:26:18.7 & -24:28:18 &    60   &  76&  76 (1)&  76&    76 &  27    &  21    &   8    &   ---           & \object{16263-2428} & YSO: \object{YLW31} Class II?\\
MMS020 & 16:27:13.0 & -24:29:55 &  1150   & 152& 152 (1)& 126&   155 &  92    &  66    &  32    & \object{B1-MM3}          &(\object{16272-2429})& sl\\
MMS021 & 16:27:33.3 & -24:27:32 &   718   & 130& 130 (1)& 110&   125 &  98    &  44    &  -9    & \object{B2-MM15}         &    n.i.    & sl\\
MMS022 & 16:26:58.6 & -24:34:06 &  4181   & 291& 291 (1)& 242&   277 & 112    &  81    &  38    & (\object{C-S})           &(\object{16269-2434})& sl\\
MMS023 & 16:27:16.1 & -24:31:13 &   186   &  83&  83 (1)&  78&    78 &  41    &  31    &  23    & \object{B1-MM4}          &(\object{16272-2430})& sl\\
MMS024 & 16:26:23.2 & -24:24:43 &   274   & 127& 127 (1)& 104&   101 &  46    &  35    &  23    & \object{LFAM3}           &     n.i.   & YSO: \object{LFAM3} Class I ?\\
MMS025 & 16:26:43.3 & -24:34:50 &   415   & 117& 117 (1)&  74&   117 & 109    &  68    &  74    & \object{WL12}            & \object{16267-2434} & YSO: \object{WL12}\\
MMS026 & 16:26:23.2 & -24:22:04 &   601   & 170& 170 (1)& 161&   176 &  72    &  28    &  25    & \object{A-MM4}           & \object{16264-2421} & sl\\
MMS027 & 16:27:15.4 & -24:30:34 &  1651   & 218& 218 (1)& 190&   207 &  73    &  72    & -39    & (\object{B1})            &(\object{16272-2430})& sl\\
MMS028 & 16:27:11.6 & -24:29:15 &   874   & 176& 176 (1)& 151&   161 &  67    &  50    &   2    & \object{B1-MM2}          &(\object{16272-2429})& sl\\
MMS029 & 16:27:28.9 & -24:26:39 &   610   & 225& 225 (1)& 208&   238 &  49    &  30    &  76    & \object{B2-MM9}/\object{12}       &(\object{16274-2427b})& sl\\
MMS030 & 16:26:08.0 & -24:20:20 &  2614   & 146& 146 (1)& 111&   153 & 142    &  92    &  10    & \object{Oph-A3}          & \object{16261-2419} & sl\\
MMS031 & 16:27:24.2 & -24:27:42 &  1318   & 194& 194 (1)& 174&   189 &  88    &  49    &  39    & \object{B2-MM3}/\object{4}/\object{5}      &(\object{16274-2427a})& sl\\
MMS032 & 16:26:40.4 & -24:27:13 &   114   &  83&  83 (1)&  72&    85 &  33    &  28    & -58    & \object{CRBR42}          & \object{16266-2427} & YSO: \object{CRBR42} Class I?\\
MMS033 & 16:27:20.1 & -24:27:09 &   624   & 138& 138 (1)& 124&   123 &  76    &  38    &  45    & \object{B2-MM2}          & \object{16273-2427} & sl\\
MMS034 & 16:26:17.0 & -24:23:40 &   126   &  70&  70 (1)&  67&    69 &  26    &  24    &  -7    & \object{CRBR12}          &    n.i.    & YSO: \object{CRBR12} Class I?\\
MMS035 & 16:27:23.8 & -24:40:54 &   921   & 167& 167 (1)& 136&   161 &  65    &  64    &  36    & \object{CRBR85} + \object{F-MM2}  &    n.i.    & YSO: Class I (?)\\
\end{tabular}
\end{sidewaystable*}
\begin{sidewaystable*}
\caption[]{List of 1.2~mm sources found in our survey.}
    \label{tab_sources1}
\begin{tabular}{lccrrrrrrrrlp{2.1cm}l}
\hline
       &  RA        &   DEC     & F$_{\rm int}$ &\multicolumn{4}{c}{F$_{\rm peak}$ (mJy)} &\multicolumn{2}{c}{FWHM (\arcsec)}&PA& MAN98  & J00 &Comment\\
       &  J2000     &   J2000   &        (mJy)&(a)& (b) &(c) &  (d)  &  maj   &  min   &        &                 &            &  \\
\hline
MMS036 & 16:27: 4.9 & -24:39:09 &  1741   & 179& 179 (1)& 162&   177 & 110    &  70    & -79    & \object{E-MM2}           & \object{16270-2439} & sl\\
MMS037 & 16:27:25.5 & -24:26:48 &  1721   & 248& 218 (2)& 224&   241 &  77    &  55    &  55    & \object{B2-MM6}          &(\object{16274-2427a})& sl\\
MMS038 & 16:27:58.6 & -24:33:34 &   595   & 135& 115 (2)& 122&   126 &  64    &  41    & -13    & ---             &    ---     & sl\\
MMS039 & 16:27: 2.0 & -24:34:46 &  1115   & 155& 138 (2)& 136&   157 &  75    &  59    & -68    & (\object{C-S})           &(\object{16269-2434})& sl\\
MMS040 & 16:27:21.1 & -24:39:57 &  1725   & 199& 177 (2)& 169&   183 &  87    &  64    & -44    & \object{F-MM1}, \object{CRBR72}   & \object{16273-2439} & YSO: Class I\\
MMS041 & 16:28:58.7 & -24:20:49 &  2827   & 164& 149 (2)& 142&   159 & 150    &  86    & -89    & ---             &    ---     & sl\\
MMS042 & 16:27:05.4 & -24:36:34 &   433   & 115&  86 (2)&  90&   105 &  68    &  43    & -62    & \object{LFAM26}          & \object{16270-2436} & YSO\\
MMS043 & 16:27:10.1 & -24:19:13 &    69   &  66&  47 (2)&  62&    74 &  29    &  24    &  11    & \object{SR21}            &     ---    & YSO: \object{SR21}\\
MMS044 & 16:26:59.1 & -24:31:26 &  2212   & 149& 135 (2)& 116&   138 & 122    &  91    & -49    & \object{C-N}             & \object{16269-2431} & sl\\
MMS045 & 16:27:27.4 & -24:39:26 &   576   & 106&  88 (2)&  88&   107 &  79    &  54    & -84    & \object{IRS44}/\object{46}; \object{CRBR88}& \object{16274-2439} & YSO: \object{IRS44}/\object{46}; \object{CRBR88}\\
MMS046 & 16:26:10.5 & -24:23:04 &   104   &  70&  52 (2)&  62&    70 &  53    &  28    &  53    & \object{A3-MM1}          & \object{16261-2423} & sl\\
MMS047 & 16:28:29.4 & -24:19:17 &   766   & 107&  96 (2)&  97&   103 &  75    &  56    &  13    & \object{D-MM1/2}         &    ---     & sl\\
MMS048 & 16:26:36.3 & -24:17:52 &   863   & 122& 107 (2)& 107&   129 &  86    &  70    & -36    & ---             & \object{16265-2418} & sl\\
MMS049 & 16:27:11.8 & -24:27:37 &   511   & 113&  94 (2)&  93&    88 &  86    &  53    &  83    & \object{B1B2-MM1}        & \object{16271-2427} & sl\\
MMS050 & 16:26:25.3 & -24:22:19 &   709   & 154& 132 (2)& 144&   153 &  78    &  39    &  19    & \object{A-MM5}           & \object{16264-2422a}& sl\\
MMS051 & 16:28:00.6 & -24:32:54 &   548   & 104&  90 (2)&  95&   105 &  74    &  46    & -34    & ---             &    ---     & sl\\
MMS052 & 16:28:32.7 & -24:17:50 &  1584   & 104&  88 (2)&  84&    96 & 118    &  83    &  31    & \object{D-MM3}/\object{4}/\object{5}       &    ---     & sl\\
MMS053 & 16:26:42.4 & -24:26:02 &   210   &  79&  64 (2)&  63&    76 &  64    &  43    & -10    & \object{A-S}             &    n.i.    & sl\\
MMS054 & 16:26:32.8 & -24:26:21 &  1479   & 136& 116 (2)& 116&   123 & 109    &  74    & -38    & n.i.            & \object{16265-2426} & sl\\
MMS055 & 16:26:48.2 & -24:33:01 &  4527   & 120& 104 (2)&  90&   111 & 180    & 140    &   3    & \object{C-W}             &    n.i.    & sl\\
MMS056 & 16:27:17.6 & -24:27:37 &   648   &  91&  80 (2)&  81&    89 &  81    &  70    & -13    & \object{B2-MM1}          & \object{16272-2427} & sl\\
MMS057 & 16:26:11.9 & -24:24:48 &   394   &  87&  71 (2)&  69&    69 &  67    &  57    & -65    & \object{A2-MM1}          & \object{16261-2424} & sl\\
MMS058 & 16:27:39.5 & -24:39:16 &    39   &  48&  29 (2)&  41&    51 &  28    &  24    & -81    & ---             & \object{16276-2439} & YSO: \object{WSB52}\\
MMS059 & 16:26:52.9 & -24:34:45 &   603   &  99&  86 (2)&  88&    82 &  72    &  62    & -17    & \object{C-MM1}           &    n.i.    & sl\\
MMS060 & 16:28:17.1 & -24:36:51 &   103   &  56&  37 (2)&  51&    53 &  39    &  26    &  42    & ---             &    ---     & YSO: Class II (?)\\
MMS061 & 16:27:19.1 & -24:28:46 &   375   &  83&  71 (2)&  75&    78 &  57    &  47    &  53    & \object{B1B2-MM2}        &    n.i.    & YSOs: \object{WL3}/\object{WL4}/\object{WL5}/\object{IRS37} Class II ...\\
MMS062 & 16:27:06.4 & -24:37:11 &   124   &  87&  56 (2)&  82&    80 &  32    &  26    &  -4    & \object{E-MM3}           & \object{16271-2437a}& YSO\\
MMS063 & 16:26:14.6 & -24:25:15 &   278   &  77&  59 (2)&  61&    70 &  58    &  49    &  34    & (\object{A2})            &(\object{16262-2425})& sl\\
MMS064 & 16:27:39.7 & -24:43:08 &   124   &  60&  39 (2)&  49&    63 &  50    &  35    & -52    & \object{IRS51}           & \object{16276-2443} & YSO: Class I \object{WLY 2-51}\\
MMS065 & 16:26:12.6 & -24:20:13 &   120   &  64&  45 (2)&  59&    51 &  42    &  34    & -54    &  n.i.           &     n.i.   & sl\\
MMS066 & 16:27:08.6 & -24:28:30 &   447   &  83&  65 (2)&  66&    70 &  85    &  43    & -11    & \object{B1-MM1}          &    n.i.    & sl\\
MMS067 & 16:27:11.2 & -24:39:26 &   227   &  63&  49 (2)&  55&    60 &  51    &  42    & -44    & \object{E-MM4}           & \object{16271-2439} & sl\\
MMS068 & 16:26:55.1 & -24:38:10 &   693   &  65&  51 (2)&  52&    54 & 145    &  71    & -81    & n.i.            &    n.i.    & sl\\
MMS069 & 16:26:17.9 & -24:25:22 &   128   &  59&  45 (2)&  49&    51 &  55    &  31    &  52    & (\object{A2})            &(\object{16262-2425})& sl\\
MMS070 & 16:27:52.7 & -24:31:43 &   340   &  67&  47 (2)&  44&    53 & 143    &  52    & -22    & ---             &    n.i.    & core + YSO \object{WLY 2-54}\\
MMS071 & 16:27:11.8 & -24:38:03 &   582   & 102&  89 (2)&  93&    99 &  94    &  56    & -57    & \object{E-MM5}; \object{WL20}     & \object{16271-2437c},  \object{16272-2438} & YSO: includes \object{WL20}\\
MMS072 & 16:27:39.1 & -24:26:57 &   337   &  71&  59 (2)&  62&    69 &  64    &  49    &  39    &   n.i.          &    n.i.    & sl\\
MMS073 & 16:27:40.2 & -24:42:34 &  1272   &  86&  75 (2)&  74&    73 & 172    &  67    & -79    &   n.i.          & \object{16276-2442}, \object{16277-2442} & sl\\
MMS074 & 16:26:23.2 & -24:20:45 &   452   &  54&  43 (2)&  44&    44 & 173    &  37    & -12    & \object{A-N} \& \object{GSS31}    & \object{16263-2419b}, \object{16263-2419a} & sl\\
\end{tabular}
\end{sidewaystable*}
\begin{sidewaystable*}
\caption[]{List of 1.2~mm sources found in our survey.}
    \label{tab_sources2}
\begin{tabular}{lccrrrrrrrrlp{2.1cm}l}
\hline
       &  RA        &   DEC     & F$_{\rm int}$ &\multicolumn{4}{c}{F$_{\rm peak}$ (mJy)} &\multicolumn{2}{c}{FWHM (\arcsec)}&PA& MAN98  & J00 &Comment\\
       &  J2000     &   J2000   &        (mJy)&(a)& (b) &(c) &  (d)  &  maj   &  min   &        &                 &            &  \\
\hline
MMS075 & 16:29:01.4 & -24:21:11 &    37   &  38&  25 (2)&  38&    45 &  42    &  18    &  14    & ---             &    ---     & sl\\
MMS076 & 16:25:56.3 & -24:20:48 &   102   &  42&  30 (2)&  34&    38 &  33    &  27    &  37    & ---             & \object{16259-2420} & YSO: TTS \object{HBC259} Class II?\\
MMS077 & 16:26:49.0 & -24:26:55 &   167   &  48&  38 (2)&  43&    40 &  77    &  41    &  61    & \object{Oph-AC2}         &    n.i.    & sl\\
MMS078 & 16:27:39.9 & -24:29:26 &    42   &  43&  24 (2)&  35&    37 &  27    &  26    &  -9    &   n.i.          &    n.i.    & sl\\
MMS079 & 16:26:33.4 & -24:18:17 &   355   &  73&  56 (2)&  60&    62 &  93    &  43    & -21    &   ---           &     n.i.   & sl\\
MMS080 & 16:27:30.5 & -24:41:56 &   779   &  77&  67 (2)&  60&    80 &  96    &  80    & -33    &   n.i.          &    n.i.    & sl\\
MMS081 & 16:27:48.3 & -24:44:44 &   588   &  66&  51 (2)&  45&    55 & 128    &  65    &  87    &   ---           & \object{16277-2444a}, \object{16277-2444b} & sl\\
MMS082 & 16:26:55.8 & -24:36:44 &   338   &  47&  34 (2)&  43&    40 & 167    &  33    &  69    & $\sim$\object{E-MM1}     &    n.i.    & sl\\
MMS083 & 16:27:38.2 & -24:27:38 &   180   &  59&  43 (2)&  50&    58 &  93    &  34    &  90    &   n.i.          &    n.i.    & sl\\
MMS084 & 16:27:37.8 & -24:46:05 &   894   &  63&  44 (2)&  40&    46 & 200    &  85    & -74    &   ---           &    n.i.    & sl\\
MMS085 & 16:26:30.4 & -24:18:39 &   406   &  69&  56 (2)&  59&    55 &  75    &  52    &  87    &   ---           &     n.i.   & sl\\
MMS086 & 16:27:19.1 & -24:31:26 &   118   &  62&  46 (2)&  56&    60 &  55    &  29    &  -7    & (\object{B1})            &(\object{16272-2430})& sl\\
MMS087 & 16:26:47.1 & -24:29:28 &  1246   &  98&  74 (3)&  87&    85 & 125    &  71    & -39    & \object{Oph-AC2}         & \object{16267-2429} & sl\\
MMS088 & 16:25:58.6 & -24:18:47 &  1786   &  88&  64 (3)&  69&    65 & 259    &  90    & -57    &    n.i.         &    n.i.    & sl\\
MMS089 & 16:27:25.6 & -24:33:32 &   640   &  56&  38 (3)&  44&    46 &  90    &  73    & -77    &   ---           &    n.i.    & sl\\
MMS090 & 16:27:01.8 & -24:26:25 &   720   &  66&  41 (3)&  46&    56 & 136    &  72    & -89    &    n.i.         &    n.i.    & sl\\
MMS091 & 16:25:58.8 & -24:31:29 &  1392   &  81&  45 (3)&  41&    41 & 178    & 146    & -29    &   ---           &    n.i.    & sl\\
MMS092 & 16:29:19.5 & -24:41:33 &   504   &  62&  29 (3)&  33&    44 &  85    &  67    & -46    &   ---           &    ---     & sl\\
MMS093 & 16:27:43.6 & -24:13:32 &  1759   &  74&  46 (3)&  44&    46 & 176    & 111    & -53    &   ---           &    ---     & sl\\
MMS094 & 16:26:59.0 & -24:22:59 &  1584   &  64&  50 (3)&  53&    52 & 222    &  94    &  67    & \object{Oph-AB}          &    n.i.    & sl\\
MMS095 & 16:27:33.3 & -24:15:40 &   992   &  56&  37 (3)&  38&    46 & 177    & 115    & -27    &   ---           &    ---     & sl\\
MMS096 & 16:26:55.4 & -24:19:14 &   545   &  51&  34 (3)&  38&    37 & 146    &  90    &  42    &   \object{Oph-G}         & \object{16269-2419} & sl\\
MMS097 & 16:28:14.2 & -25:02:58 &   589   &  70&  29 (3)&  41&    31 & 138    &  69    & -19    &   ---           &    ---     & sl\\
MMS098 & 16:25:47.3 & -24:29:54 &   573   &  69&  26 (3)&  34&    25 & 201    &  85    &  66    &   ---           &    n.i.    & sl\\
MMS099 & 16:28:07.4 & -24:35:20 &  2658   &  64&  46 (3)&  45&    55 & 283    & 125    & -64    &   ---           &    ---     & sl\\
MMS100 & 16:27:56.4 & -24:42:48 &   637   &  47&  28 (3)&  36&    34 & 212    &  64    &  82    &   ---           &    ---     & sl\\
MMS101 & 16:27:12.7 & -24:21:23 &   648   &  49&  32 (3)&  34&    34 & 127    &  96    & -88    &   ---           &    ---     & sl\\
MMS102 & 16:29:07.7 & -24:35:03 &   324   &  50&  21 (3)&  41&    25 & 135    &  48    & -88    &   ---           &    ---     & sl\\
MMS103 & 16:27:55.2 & -24:35:07 &   323   &  50&  29 (3)&  44&    38 &  86    &  51    & -75    &   ---           &    ---     & sl\\
MMS104 & 16:26:27.9 & -24:33:36 &  1166   &  65&  41 (3)&  42&    46 & 141    & 119    & -54    &  ---            &    n.i.    & sl\\
MMS105 & 16:25:44.6 & -24:16:56 &  1068   &  55&  32 (3)&  32&    31 & 191    & 131    &  49    &  ---            &    n.i.    & sl\\
MMS106 & 16:26:48.9 & -24:23:33 &   169   &  39&  25 (3)&  36&    38 &  64    &  50    &  49    &  n.i.           &    n.i.    & sl\\
MMS107 & 16:27:40.3 & -24:35:50 &  1488   &  57&  39 (3)&  44&    40 & 277    & 104    & -43    &  ---            &    n.i.    & sl\\
MMS108 & 16:27:20.6 & -24:24:03 &   521   &  50&  32 (3)&  42&    42 & 123    &  64    & -19    & (\object{B3})            &    n.i.    & sl\\
MMS109 & 16:29:11.2 & -24:40:58 &   590   &  47&  28 (3)&  33&    28 & 157    &  98    & -75    &  ---            &    ---     & sl\\
MMS110 & 16:25:49.4 & -24:09:34 &   255   &  40&  20 (3)&  26&    26 & 176    &  64    &  66    &  ---            &    ---     & sl\\
MMS111 & 16:27:26.1 & -24:36:21 &   180   &  34&  19 (3)&  29&    26 & 111    &  53    &  68    &  ---            &    n.i.    & sl\\
MMS112 & 16:27:10.7 & -24:26:15 &   273   &  49&  28 (3)&  40&    38 &  95    &  50    &  28    &  n.i.           &    n.i.    & sl\\
MMS113 & 16:26:34.3 & -24:35:27 &   341   &  51&  24 (3)&  32&    26 & 115    &  68    & -58    &  ---            &    n.i.    & sl\\
MMS114 & 16:27:28.0 & -24:20:32 &   223   &  40&  21 (3)&  29&    26 &  88    &  58    &  59    &  ---            &    ---     & sl\\
MMS115 & 16:26:16.2 & -24:31:21 &   757   &  58&  30 (3)&  31&    31 & 158    & 109    & -76    &  ---            &    n.i.    & sl\\
\end{tabular}
\end{sidewaystable*}
\begin{sidewaystable*}
\caption[]{List of 1.2~mm sources found in our survey.}
    \label{tab_sources3}
\begin{tabular}{lccrrrrrrrrlp{2.1cm}l}
\hline
       &  RA        &   DEC     & F$_{\rm int}$ &\multicolumn{4}{c}{F$_{\rm peak}$ (mJy)} &\multicolumn{2}{c}{FWHM (\arcsec)}&PA& MAN98  & J00 &Comment\\
       &  J2000     &   J2000   &        (mJy)&(a)& (b) &(c) &  (d)  &  maj   &  min   &        &                 &            &  \\
\hline
MMS116 & 16:29:01.7 & -25:06:30 &  1214   &  76&  37 (3)&  38&    39 & 181    & 132    & -67    &  ---            &    ---     & sl\\
MMS117 & 16:27:39.1 & -24:41:15 &   315   &  49&  31 (3)&  43&    43 &  71    &  54    &  82    &  n.i.           &    n.i.    & sl\\
MMS118 & 16:27:15.1 & -24:24:12 &   177   &  38&  22 (3)&  33&    30 &  77    &  54    &  89    & (\object{B3})            &    n.i.    & sl\\
MMS119 & 16:29:18.9 & -24:35:46 &   248   &  50&  20 (3)&  24&    27 &  80    &  73    &  17    &  ---            &    ---     & sl\\
MMS120 & 16:28:54.3 & -24:59:08 &    94   &  33&  14 (3)&  27&    24 &  78    &  36    & -54    &   ---           &    ---     & sl\\
MMS121 & 16:26:04.1 & -24:35:59 &   185   &  41&  18 (3)&  25&    23 &  88    &  65    & -78    &   ---           &    n.i.    & sl\\
MMS122 & 16:28:59.6 & -24:41:08 &   244   &  43&  20 (3)&  27&    24 &  91    &  75    &  24    &   ---           &    ---     & sl\\
MMS123 & 16:26:17.3 & -24:16:57 &   273   &  53&  24 (3)&  24&    29 & 138    &  99    &  73    &   ---           &    n.i.    & sl\\
MMS124 & 16:26:37.8 & -24:13:51 &   113   &  29&  17 (3)&  22&    23 &  95    &  48    &  54    &   ---           &    ---     & sl\\
MMS125 & 16:27:27.2 & -24:16:56 &   209   &  37&  21 (3)&  29&    28 &  96    &  52    & -24    &   ---           &    ---     & sl\\
MMS126 & 16:28:21.7 & -24:36:21 &   411   & 121& 121 (1)& 102&   118 &  53    &  43    &  21    &   ---           &    ---     &YSO: \object{IRAS 16253-2429}, Class 0?\\
MMS127 & 16:26:44.4 & -24:12:25 &   448   &  42&  21 (3)&  21&    20 & 125    &  89    & -66    &   ---           &    ---     & sl\\
MMS128 & 16:29:17.8 & -24:10:49 &   977   &  64&  22 (4)&  30&    30 & 148    & 125    &  10    &   ---           &    ---     & sl\\
MMS129 & 16:29:08.8 & -24:13:55 &  1165   &  75&  23 (4)&  24&    28 & 192    & 163    & -67    &   ---           &    ---     & sl\\
MMS130 & 16:28:34.4 & -24:36:28 &   823   &  44&  20 (4)&  28&    26 & 233    &  98    &  89    &   ---           &    ---     & sl\\
MMS131 & 16:28:33.4 & -25:03:19 &   958   &  48&  20 (4)&  25&    22 & 238    & 142    &  30    &   ---           &    ---     & sl\\
MMS132 & 16:27:00.1 & -25:00:22 &   952   &  70&  16 (4)&  28&    18 & 306    & 123    & -53    &   ---           &    ---     & sl\\
MMS133 & 16:29:05.9 & -24:58:37 &   879   &  44&  15 (4)&  20&    15 & 307    & 159    &  65    &   ---           &    ---     & sl\\
MMS134 & 16:25:33.6 & -24:35:03 &   426   &  49&  13 (4)&  17&    17 & 161    & 124    &  43    &   ---           &    n.i.    & sl\\
MMS135 & 16:27:12.9 & -24:50:02 &   234   &  32&  10 (4)&  19&    15 & 148    &  86    & -51    &   ---           &    ---     & sl\\
MMS136 & 16:27:53.2 & -24:26:13 &   531   &  37&  12 (4)&  19&    17 & 168    & 103    &  35    &   ---           &    ---     & sl\\
MMS137 & 16:27:57.7 & -24:11:00 &   340   &  40&  11 (4)&  21&    15 & 158    &  90    & -83    &   ---           &    ---     & sl\\
MMS138 & 16:28:58.5 & -24:46:59 &   398   &  45&  11 (4)&  18&    14 & 153    & 118    &  31    &   ---           &    ---     & sl\\
MMS139 & 16:28:07.2 & -24:59:55 &   770   &  53&  12 (4)&  17&    16 & 168    & 122    & -28    &   ---           &    ---     & sl\\
       &            &           &         &    &        &    &       &        &        &        &                 &            &\\
MMS140 & 16:27:06.5 & -24:38:10 &   144   &  62&        &  54&    54 &  55    &  26    & -71    & \object{WL17}            & \object{16271-2438} &YSO: \object{WL17}\\
MMS141 & 16:25:38.0 & -24:22:32 &    36   &  40&        &  40&    48 &  26    &  21    &  -9    &   ---           & \object{16256-2422} &YSO: Class I/II?\\
MMS142 & 16:27:17.2 & -24:40:58 &  1058   &  83&        &  68&    67 & 133    &  66    & -58    &  n.i.           &    n.i.    & sl\\
MMS143 & 16:27:27.6 & -24:24:38 &   342   &  53&        &  42&    36 & 100    &  51    & -53    & (\object{B3})            &    n.i.    & sl\\
\end{tabular}
\end{sidewaystable*}

\noindent{}{\bf MMS001:}
ridge NE of \object{VLA1623}: includes starless cores \object{SM1} and \object{SM1N}; no optical/IR
stellar counterpart.

\noindent{}{\bf MMS002:}
\object{VLA1623} Class~0 \citep{1993ApJ...406..122A}; no opt/ir counterpart.

\noindent{}{\bf MMS003:}
pointlike source; optical/IR stellar counterpart \object{YLW32}.

\noindent{}{\bf MMS004:}
\object{SM2} starless core; no opt/ir counterpart

\noindent{}{\bf MMS005:}
compact condensation in filament W of \object{SM1}/\object{VLA1623} ridge; YSO \object{LFAM1}; no
optical counterpart; bright NIR source, extended K-band nebula.

\noindent{}{\bf MMS006:}
pointlike source; \object{GSS39}/\object{Elia2-27}; no opt.\ coutnerpart; IR counterpart.

\noindent{}{\bf MMS007:}
part of filament extending NE of \object{SM1}/\object{VLA1623}; no clear opt/IR stellar
counterpart.

\noindent{}{\bf MMS008:}
compact source: multiple T Tauri star \object{SR24}; the flux seems to be associated
with the (so far) unresolved southern single component, as claimed by
\citet{1998A&A...330..549N} and confirmed recently by 
\citet{2005ApJ...619L.175A}, rather than the northern 
(close binary) component. Our flux measurement is in good agreement
with the result of \citet{1998A&A...330..549N}.

\noindent{}{\bf MMS009:}
compact source; no opt.\ stellar counterpart; NIR star \object{Elias 2-29} (also 
includes \object{GY210}).

\noindent{}{\bf MMS010:}
compact source; optical/IR counterpart  \object{YLW34}

\noindent{}{\bf MMS011:}
compact source; YSO \object{WLY 2-43}; Class~I; no opt.\ stellar counterpart; 
NIR star; high extinction.

\noindent{}{\bf MMS012:}
compact, but clearly extended source; \object{SMM16267} of \citet{1999ApJ...513L.139W}
(suggesting it to be a prestellar core)
no optical counterpart; high extinction; fuzzy K band emission south of clump

\noindent{}{\bf MMS013:}
compact source; no opt.\ star; NIR star  \object{GSS26}.

\noindent{}{\bf MMS014:}
elongated feature SE of MMS005; no optical counterpart; some IR stars might
be associated. Breaks up into chain of three subcores in MAN98.

\noindent{}{\bf MMS015:}
bright compact core; no opt.\ stellar counterpart; NIR counterpart \object{VSSG17}; YSO; high opt.\ extinction.

\noindent{}{\bf MMS016:}
bright elongated core; YSO \object{VSSG18} appears to be associated with this feature.

\noindent{}{\bf MMS017:}
part of filament extending NE of \object{SM1}/\object{VLA1623}; very faint K-band source
at NE tip?

\noindent{}{\bf MMS018:}
compact core; no opt./NIR stellar counterpart.

\noindent{}{\bf MMS019:}
faint compact source; faint opt./bright NIR stellar counterpart \object{YLW31}

\noindent{}{\bf MMS020:}
bright core; no clear opt./NIR stellar counterpart; high opt.\ extinction.

\noindent{}{\bf MMS021:}
highly elongated; no opt./NIR stellar counterpart.

\noindent{}{\bf MMS022:}
bright large core; no opt./NIR stellar counterpart; high extinction.

\noindent{}{\bf MMS023:}
compact source; no clear optical counterpart; possibly IR star associated;
high opt.\ extinction.

\noindent{}{\bf MMS024:}
\object{LFAM3} faint opt., bright nebulous NIR source; visible on J00 map, but not
noted as separate feature

\noindent{}{\bf MMS025:}
compact source (possibly surrounded by extended Halo); no opt.\ stellar 
counterpart; IR stellar counterpart \object{WL12}; high opt.\ extinction.

\noindent{}{\bf MMS026:}
elongated core NE of MMS005; no optical counterpart; faint K-band source?

\noindent{}{\bf MMS027:}
bright core; no clear opt./NIR stellar counterpart; possibly IR star to the SE associated; high opt.\ extinction.

\noindent{}{\bf MMS028:}
compact core; no clear opt./NIR stellar counterpart; high opt.\ extinction.

\noindent{}{\bf MMS029:}
elongated core. No opt./NIR stellar counterpart.

\noindent{}{\bf MMS030:}
extended core at SE tip of ridge (MMS088); no related optical feature.

\noindent{}{\bf MMS031:}
elongated core; no opt./NIR stellar counterpart.

\noindent{}{\bf MMS032:}
faint compact source; no optical counterpart; faint IR star \object{CRBR42};
\object{ISO-Oph 54}; \object{[GY92] 91} 

\noindent{}{\bf MMS033:}
elongated core; no opt./NIR stellar counterpart.

\noindent{}{\bf MMS034:}
faint compact source; no optical counterpart; IR stellar counterpart \object{CRBR12}.

\noindent{}{\bf MMS035:}
compact source; no opt.\ stellar counterpart; NIR stellar counterpart, 
YSO \object{CRBR85} Class~I?; splits into two cores in MAN98 (YSO \object{CRBR85} and \object{F-MM2}
starless core).

\noindent{}{\bf MMS036:}
no stellar counterpart for main core; MAN98 \object{E-MM2b} might have faint
K-band counterpart.

\noindent{}{\bf MMS037:}
elongated core; no opt./NIR stellar counterpart.

\noindent{}{\bf MMS038:}
bright compact core; no opt./NIR stellar counterpart; high extinction.

\noindent{}{\bf MMS039:}
bright core adjacent to MMS022; no opt./NIR stellar counterpart;
high extinction.

\noindent{}{\bf MMS040:}
elongated core; no opt./NIR stellar counterpart; high extinction.
Contains MAN98 starless core \object{F-MM1} and, at its north-western tip, the
YSO \object{CRBR72} (Class~I).

\noindent{}{\bf MMS041:}
bright core; no opt./NIR stellar counterpart; near tip of cometary shaped
high-optical-extinction cloud. The edge of the cloud are outlined by faint
extensions to the east and west of this core.

\noindent{}{\bf MMS042:}
compact source; YSO \object{LFAM26}; no opt.\ stellar counterpart; faint K-Band source;
optical extinction ridge.

\noindent{}{\bf MMS043:}
compact faint source, YSO \object{SR21}

\noindent{}{\bf MMS044:}
large core; compact source at center? no opt./NIR stellar counterpart; high extinction.

\noindent{}{\bf MMS045:}
faint compact core; no opt. stars; 3 NIR YSO stellar sources: \object{IRS44}/\object{IRS46};
\object{CRBR88}; 3 subsources det. by MAN98

\noindent{}{\bf MMS046:}
faint compact source; no clear opt/NIR counterpart.

\noindent{}{\bf MMS047:}
bright core; no opt./NIR stellar counterpart; at tip of cometary dark cloud.

\noindent{}{\bf MMS048:}
small core; no optical/NIR counterpart; high extinction.
Part of small clump (together with MMS079 and MMS085).

\noindent{}{\bf MMS049:}
small faint core; no opt./NIR stellar counterpart; high opt.\ extinction.

\noindent{}{\bf MMS050:}
core in SM1 ridge; no clear opt/NIR stellar counterpart.

\noindent{}{\bf MMS051:}
compact core; no opt./NIR stellar counterpart; high extinction.

\noindent{}{\bf MMS052:}
rather bright elongated core; no opt./NIR stellar counterpart; tail of  cometary dark cloud.

\noindent{}{\bf MMS053:}
small faint core; high optical extinction; no stellar opt./NIR counterpart.

\noindent{}{\bf MMS054:}
faint core; high optical extinction; no clear opt/IR stellar counterpart.

\noindent{}{\bf MMS055:}
faint core; no clear opt./NIR stars associated; high extinction.

\noindent{}{\bf MMS056:}
diffuse core; no opt./NIR stellar counterpart; high opt.\ extinction.

\noindent{}{\bf MMS057:}
small core; no clear optical/NIR counterpart. 
Part of small core (together with MMS063 and MMS069).

\noindent{}{\bf MMS058:}
faint compact source; YSO \object{WSB52}/\object{LFAM p8}/\object{[GY92] 314}; opt./NIR star.

\noindent{}{\bf MMS059:}
no opt./NIR stellar counterpart; high opt.\ extinction.

\noindent{}{\bf MMS060:}
faint compact source; opt/NIR stellar counterpart \object{WSB~60} = 
\object{BKLT~J162816$-$243657}.

\noindent{}{\bf MMS061:}
elongated core; superposition of several NIR YSOs: \object{WL3}/\object{WL4}/\object{WL5}/\object{IRS37}; no 
optical stellar counterpart; high extinction. Class~II sources. MAN98 classify
this feature as composite pre-collapse core, but it might actually just be a
superposition of circumstellar dust emission from the 4 NIR YSOs visible in 
this area.

\noindent{}{\bf MMS062:}
compact source; associated with edge-on disk YSO \citep{2000A&A...364L..13B}.

\noindent{}{\bf MMS063:}
small core; no clear optical/NIR counterpart.
Part of small clump (together with MMS057 and MMS069).

\noindent{}{\bf MMS064:}
faint compact source; no opt.\ star; NIR stellar counterpart; 
YSO: \object{IRS51} = \object{WLY 2-51} = \object{YLW 45}.

\noindent{}{\bf MMS065:}
faint compact source; seems to be associated with IR RN; YSO?
hardly visible on MAN98 map; visible on J00 map, but not noted as 
separate feature.

\noindent{}{\bf MMS066:}
faint elongated core; no opt./NIR stellar counterpart; high opt.\ extinction.

\noindent{}{\bf MMS067:}
no opt./NIR stellar counterpart.

\noindent{}{\bf MMS068:}
extended core; no clear opt./IR stellar conterpart. There is emission
visible in the MAN98 map, but
there is some morphological difference from our and the J00 maps. 

\noindent{}{\bf MMS069:}
small core; no clear optical/NIR counterpart.
Part of small clump (together with MMS057 and MMS063).

\noindent{}{\bf MMS070:}
faint core; core + envelope structure? No opt.\ stellar counterpart; 
NIR star \object{WLY 2-54}.

\noindent{}{\bf MMS071:}
elongated core; no opt./NIR stellar counterpart; high extinction. Might split
into a chain of faint sources. Includes NIR YSO \object{WL20} at its south-eastern tip.

\noindent{}{\bf MMS072:}
faint core at eastern end of \object{clump B2}; no opt./NIR stellar counterpart.

\noindent{}{\bf MMS073:}
elongated core; no direct stellar counterpart, but \object{[GY92] 301} is within
core area.

\noindent{}{\bf MMS074:}
filament extending north of \object{SM1} core; \object{GSS31} on base of filament; 
possibly very faint nebulosity at tip.

\noindent{}{\bf MMS075:}
faint compact source nearby MMS041; no opt./NIR stellar counterpart.

\noindent{}{\bf MMS076:}
faint compact source; bright opt/NIR counterpart; \object{V* V2058 Oph}; \object{HBC 259}; \object{SR4}

\noindent{}{\bf MMS077:}
small faint core; located on high-extinction filament; no IR counterpart;
not clearly visible in MAN98 map, part of their clump \object{Oph-AC2}

\noindent{}{\bf MMS078:}
faint small core, no opt./NIR counterpart.

\noindent{}{\bf MMS079:}
small core; no optical/NIR counterpart; high extinction.
Part of small clump (together with MMS048 and MMS085).

\noindent{}{\bf MMS080:}
faint large core; no opt./NIR stellar counterpart.

\noindent{}{\bf MMS081:}
large, elongated core; 3 NIR (YSO?) stars on our around the core: 
\object{GY344}, \object{GY370}, \object{GY352}; high opt.\ extinction.

\noindent{}{\bf MMS082:}
faint filament; no opt/NIR stellar counterpart; high extinction.
Corresponds roughly to MAN98 \object{E-MM1}, but differs in morphology from our and
the J00 maps.

\noindent{}{\bf MMS083:}
faint core at eastern end of \object{clump B2}; no opt./NIR stellar counterpart.

\noindent{}{\bf MMS084:}
large core; no clear opt./NIR stellar counterpart; close to edge of
high-extinction cloud; close to edge of J00 map; there is some emission,
but it is not marked as separate feature.

\noindent{}{\bf MMS085:}
small core; no optical/NIR counterpart; high extinction.
Part of small clump (together with MMS048 and MMS079).

\noindent{}{\bf MMS086:}
compact source; no clear opt./NIR counterpart; high opt.\ extinction.

\noindent{}{\bf MMS087:}
large core; no opt./NIR stellar counterpart; high extinction patch.

\noindent{}{\bf MMS088:}
SE-NW elongated ridge bounded by MMS030 and MMS105; apparently illuminated
from the SW by \object{HD147889} (\object{SR1})

\noindent{}{\bf MMS089:}
faint extended core; no clear opt./NIR stellar counterpart; some bright NIR
stars around; extinction patch.

\noindent{}{\bf MMS090:}
faint core; on optical extinction filament; 
no clear opt/IR stellar counterpart.

\noindent{}{\bf MMS091:}
faint large core; no stellar counterpart; optical extinction patch.

\noindent{}{\bf MMS092:}
faint core, no opt./NIR stellar counterpart; MMS092, MMS109, and MMS122 form
a filament associated with high extinction.

\noindent{}{\bf MMS093:}
low surface brightness core; on high optical extinction lane 
(along with MMS095, MMS125, and MMS137); no clear opt/NIR stellar counterpart.

\noindent{}{\bf MMS094:}
part of larger ridge; no opt./IR stellar counterpart; on opt.\ extinction lane.

\noindent{}{\bf MMS095:}
low surface brightness core; on high optical extinction lane
(along with MMS093, MMS125, and MMS137); no clear opt/NIR stellar counterpart.

\noindent{}{\bf MMS096:}
small low surface brightness core; no clearly associated optical/NIR 
counterpart.

\noindent{}{\bf MMS097:}
extended low surface brightness feature; no clear feature seen on DSS/2MASS.

\noindent{}{\bf MMS098:}
faint core; no stellar counterpart; optical extinction patch.

\noindent{}{\bf MMS099:}
ridgelike, elongated core; no opt./NIR stellar counterpart; high-extinction
lane.

\noindent{}{\bf MMS100:}
large elongated core; no opt./NIR stellar counterpart; high extinction.

\noindent{}{\bf MMS101:}
part of larger ridge; on high optical extinction lane; no clear opt/NIR
stellar counterpart; on edge of J00 map.

\noindent{}{\bf MMS102:}
very faint large core; no opt./NIR stellar counterpart; extinction patch.

\noindent{}{\bf MMS103:}
faint core;  no opt./NIR stellar counterpart; high extinction.

\noindent{}{\bf MMS104:}
faint large core; no opt/NIR stellar counterpart; high-extinction.

\noindent{}{\bf MMS105:}
extended low surface brightness  feature at NW end of ridge (MMS088);
associated with diffuse K-band RN at SW edge; maybe some optical extiction
patches associated.

\noindent{}{\bf MMS106:}
core on larger ridge; no opt./IR stellar counterpart; on opt. extinction lane.

\noindent{}{\bf MMS107:}
faint core; no clear opt./NIR stellar counterpart, but there are some IR
stars in the core area which might belong to this feature; optical
extinction filament.

\noindent{}{\bf MMS108/MMS118/MMS143:}
faint cores; no opt./NIR stellar counterpart; high optical extinction.
part of MAN98 \object{B3}.

\noindent{}{\bf MMS109:}
faint core, no opt./NIR stellar counterpart; MMS092, MMS109, and MMS122 form
a filament associated with high extinction.

\noindent{}{\bf MMS110:}
small faint core; small optical extinction whisps; IR star near peak position.

\noindent{}{\bf MMS111:}
faint core; no opt./NIR stellar counterpart; high extinction.

\noindent{}{\bf MMS112:}
faint core; no clear opt./NIR stellar counterpart; high opt.\ extinction lane?

\noindent{}{\bf MMS113:}
faint large core; possibly faint compact source also; no opt./NIR stellar
counterpart; high-extinction lane.

\noindent{}{\bf MMS114:}
faint core; located on high optical extinction filament; no clear opt./NIR stellar counterpart.

\noindent{}{\bf MMS115:}
faint, large core; no opt./NIR stellar counterpart; high extinction lane.

\noindent{}{\bf MMS116/120:}
extended low surface brightness features; no clear features seen on DSS/2MASS.

\noindent{}{\bf MMS117:}
faint large core; no opt./NIR stellar counterpart; high extinction.

\noindent{}{\bf MMS118:} see MMS108

\noindent{}{\bf MMS119:}
very faint large core; no opt./NIR stellar counterpart; extinction patch.

\noindent{}{\bf MMS120:} see MMS116

\noindent{}{\bf MMS121:} faint core; no stellar counterpart, roughly corresponds to
optical extinction patch.

\noindent{}{\bf MMS122:}
faint core, no opt./NIR stellar counterpart; MMS092, MMS109, and MMS122 form
a filament associated with high extinction.

\noindent{}{\bf MMS123:}
faint core; no optical counterpart; NIR star within core area;
high extinction. Faint on J00 map.

\noindent{}{\bf MMS124:}
faint core, no opt./NIR stellar counterpart; high extinction.

\noindent{}{\bf MMS125:}
faint core; on high optical extinction lane
(along with MMS093, MMS095, and MMS137); no clear opt/NIR stellar counterpart.

\noindent{}{\bf MMS126:}
compact source; no opt./NIR stellar counterpart; \object{IRAS~16253$-$2429}? Class~0?
Main source might have a companion to the north-east.

\noindent{}{\bf MMS127:}
faint core; the YSO \object{YLW37} (\object{BKLT J162646-241203}) is located at its 
north-eastern periphery.

\noindent{}{\bf MMS128:}
very faint large core; no opt./NIR stellar counterpart; extinction patch.

\noindent{}{\bf MMS129:}
very faint large core; no opt./NIR stellar counterpart; extinction patch.

\noindent{}{\bf MMS130:}
large, faint, elongated core; no opt./NIR star associated; extinction lane.

\noindent{}{\bf MMS131/132/133/139:}
extended low surface brightness features; no clear features seen on DSS/2MASS.

\noindent{}{\bf MMS134:}
low surface brightness; no opt.\ stellar counterpart, diffuse nebulosity;
NIR star \object{GSS15}/\object{BKLT J162535-243400} at north-eastern periphery. Faint stuff
on J00 map.

\noindent{}{\bf MMS135:}
low surface brightness; small optical extinction patch.

\noindent{}{\bf MMS136:}
low surface brightness; 2 optical/NIR stars, \object{V 2059 Oph} and \object{ISO-Oph 180}
\citep{2001A&A...372..173B} within core area. On edge of MAN98 and J00 maps.

\noindent{}{\bf MMS137:}
low surface brightness; on high optical extinction lane
(along with MMS093, MMS095, and MMS125); no clear opt/NIR stellar counterpart.

\noindent{}{\bf MMS138:}
low surface brightness; some opt./NIR stars visible at core periphery.

\noindent{}{\bf MMS139:} see MMS131

\noindent{}{\bf MMS140:}
faint compact source plus more extended core; no opt.\ counterpart; NIR
star \object{WL17}; high extinction.

\noindent{}{\bf MMS141:}
faint compact source; stellar NIR counterpart; optically very faint; 
Class I/II?

\noindent{}{\bf MMS142:}
faint extended core.

\noindent{}{\bf MMS143:} see MMS108

\section{Orientation        
\label{chap_orient}}        
 
On the largest scale, the $\rho$ Ophiuchi cloud possesses southeast-northwest 
ridges (general orientation of $\sim$ 145$^\circ$) along with filamentary 
tails stretching in the NNE direction ($\sim$ 70$^\circ$) over four degrees
on the sky \citep{1989ApJ...338..902L}.

\begin{figure}[th]
\centerline{\psfig{file=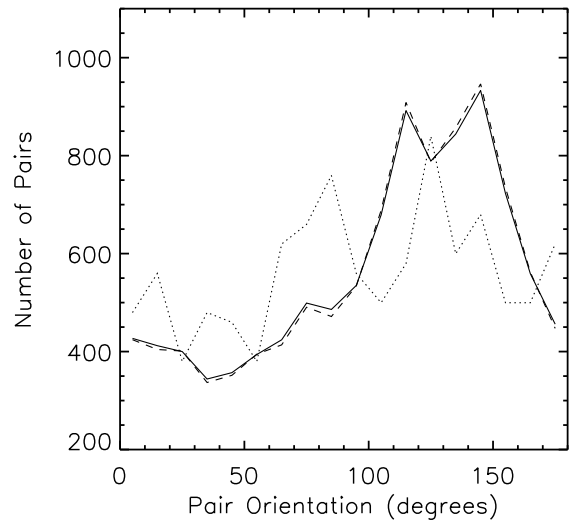,width=8.5cm}}
\caption{The number distribution of core-core orientations determined for each
   pair of cores (a total of 10,153 pairs; solid line), each pair separated
   by less than 30,000\,AU (508 pairs; dotted line) and those separated by
   more than 30,000\,AU (dashed line). The latter two distributions have been
   normalised so that the total number is constant (i.e. displaying equal areas
   under the lines).}
\label{pairorient}
\end{figure}

A preferred cloud direction on sub-degree scales is also apparent, as shown
on our 1.2\,mm image (Fig.\,\ref{fig_mosaic}). Here, we quantify this by 
determining the orientation between each pair of cores. The resulting number
distribution is displayed in Fig.\,\ref{pairorient}.

Not surprisingly, the result is that the preferred core-pair direction is not only
maintained on the sub-degree scale (solid line) but remains just as
prominent below the separation of 30,000\,AU (0.06$^\circ$). Thus the preferred
direction of $\sim$ 130 - 150$^\circ$ corresponds to the ridge direction --
the expected direction of compression from shocks transmitted from the \object{Sco OB2 
association} \citep{1977AJ.....82..198V,1989ApJ...338..925L}.

\begin{figure}[th]
\centerline{\psfig{file=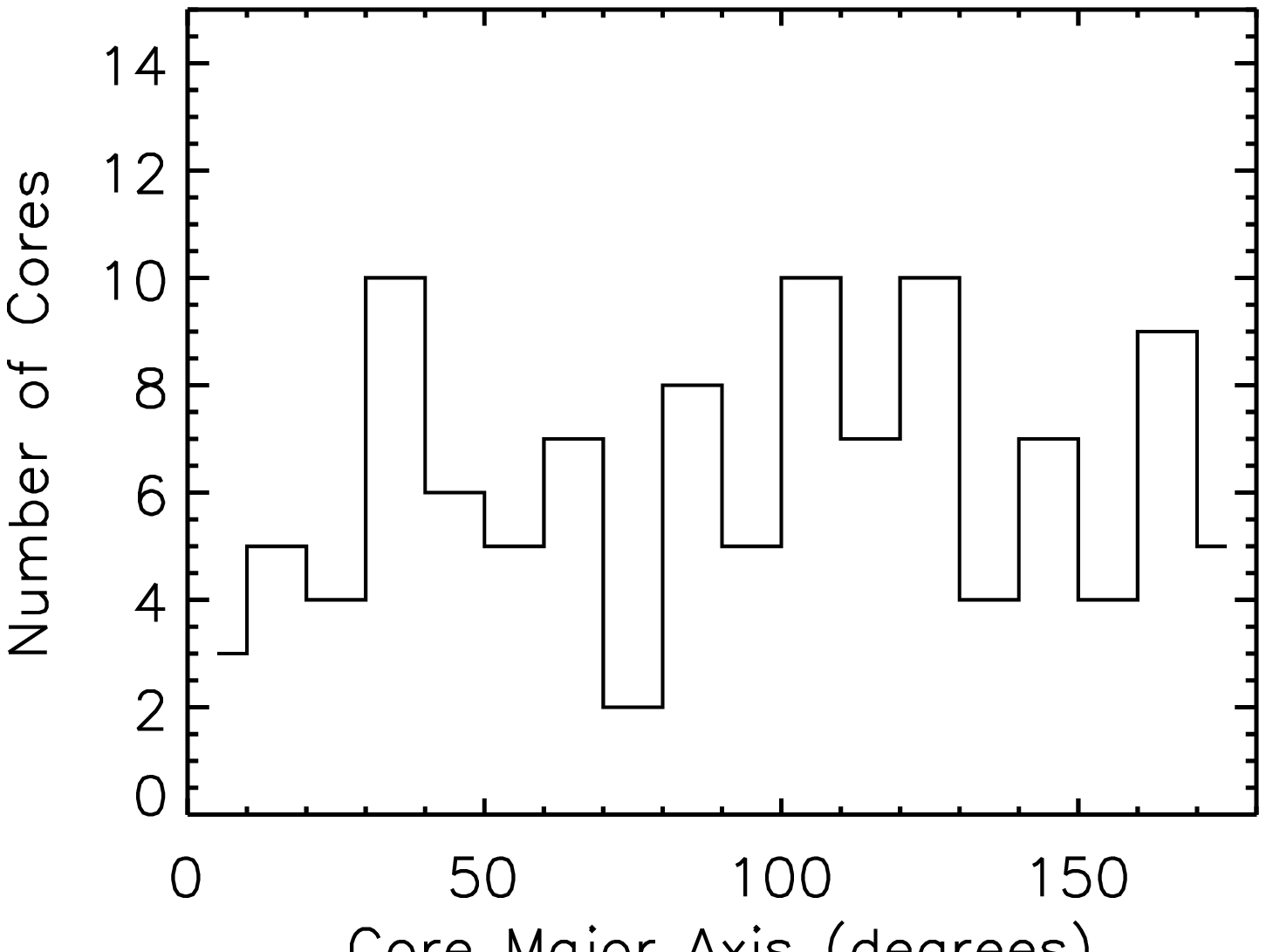,width=8.5cm}}
\centerline{\psfig{file=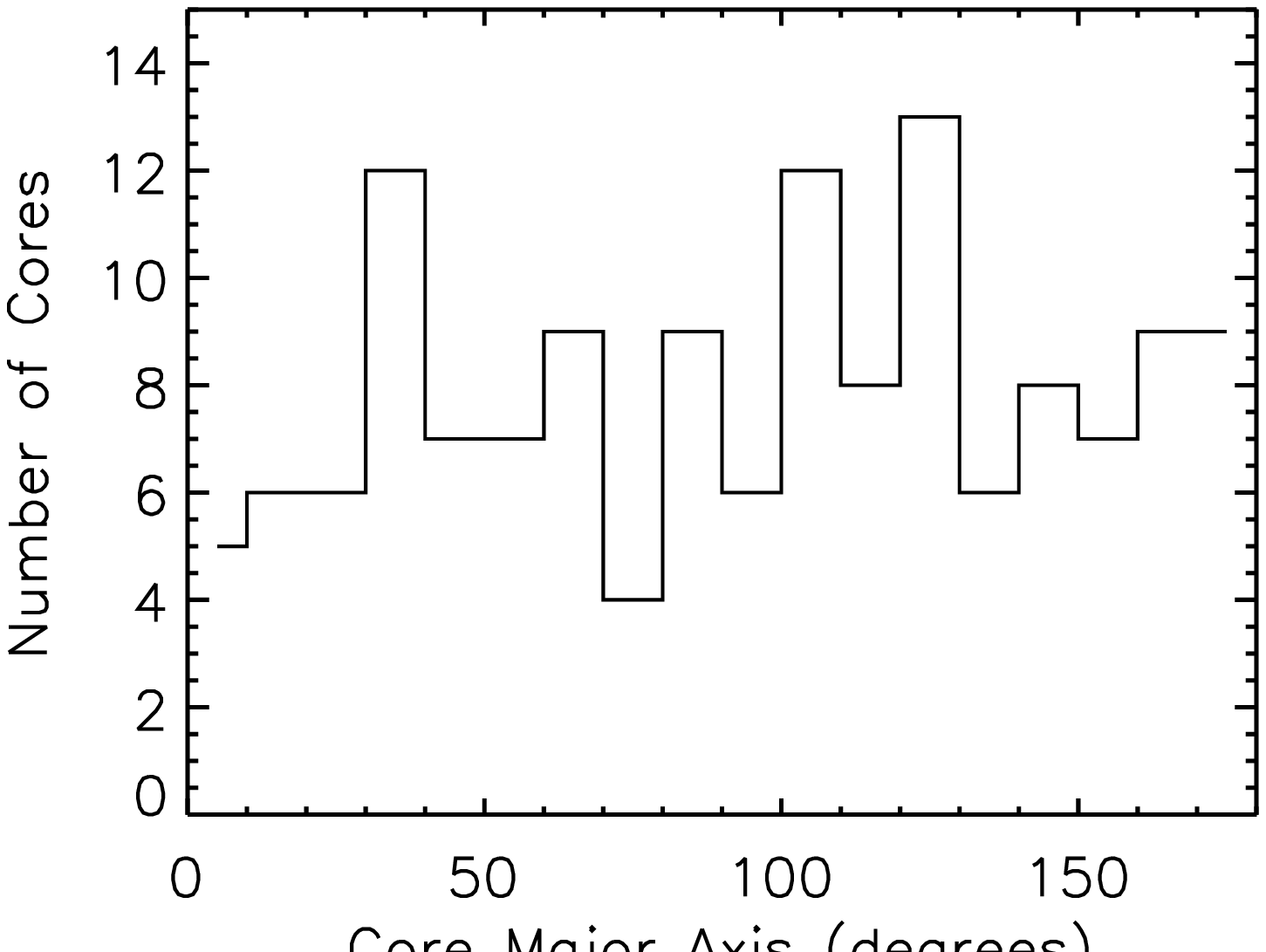,width=8.5cm}}
\caption{The number distribution of orientations of the major axes of the 
   111 starless cores (upper panel) and the entire 143 cores (lower panel). }
\label{majoraxes}
\end{figure}

Is the preferred direction still present on even smaller scales? To answer 
this, we plot the number distribution of the orientations of the major axis
of each core in Fig.\,\ref{majoraxes}. We find no obvious trend in the data
although the statistics are quite small. This result applies to the entire
sample as well as the starless cores which implies that any preferred 
direction is lost at any early stage in core development.

\begin{figure}[th]
\centerline{\psfig{file=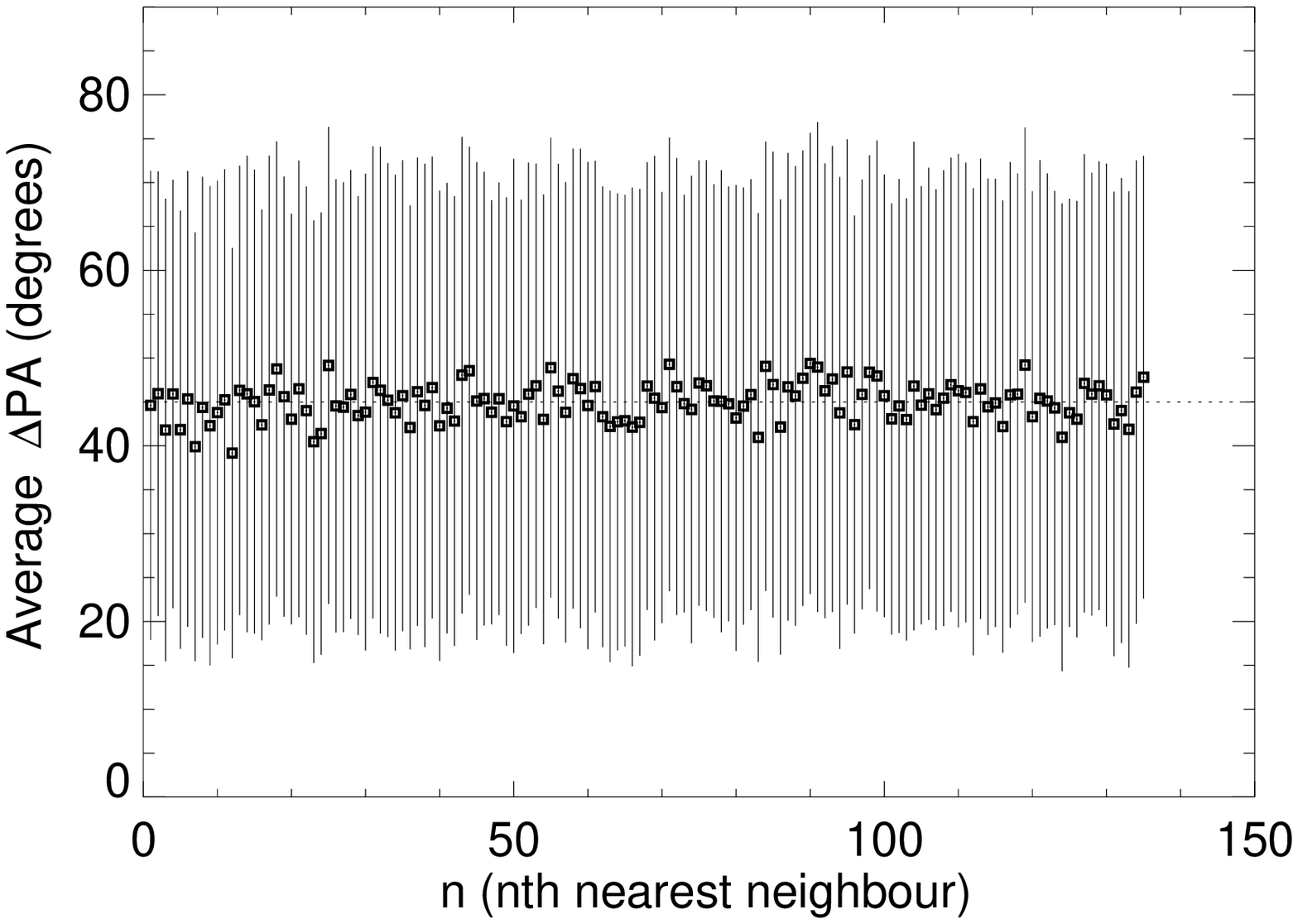,width=8.5cm,angle=0}}
\centerline{\psfig{file=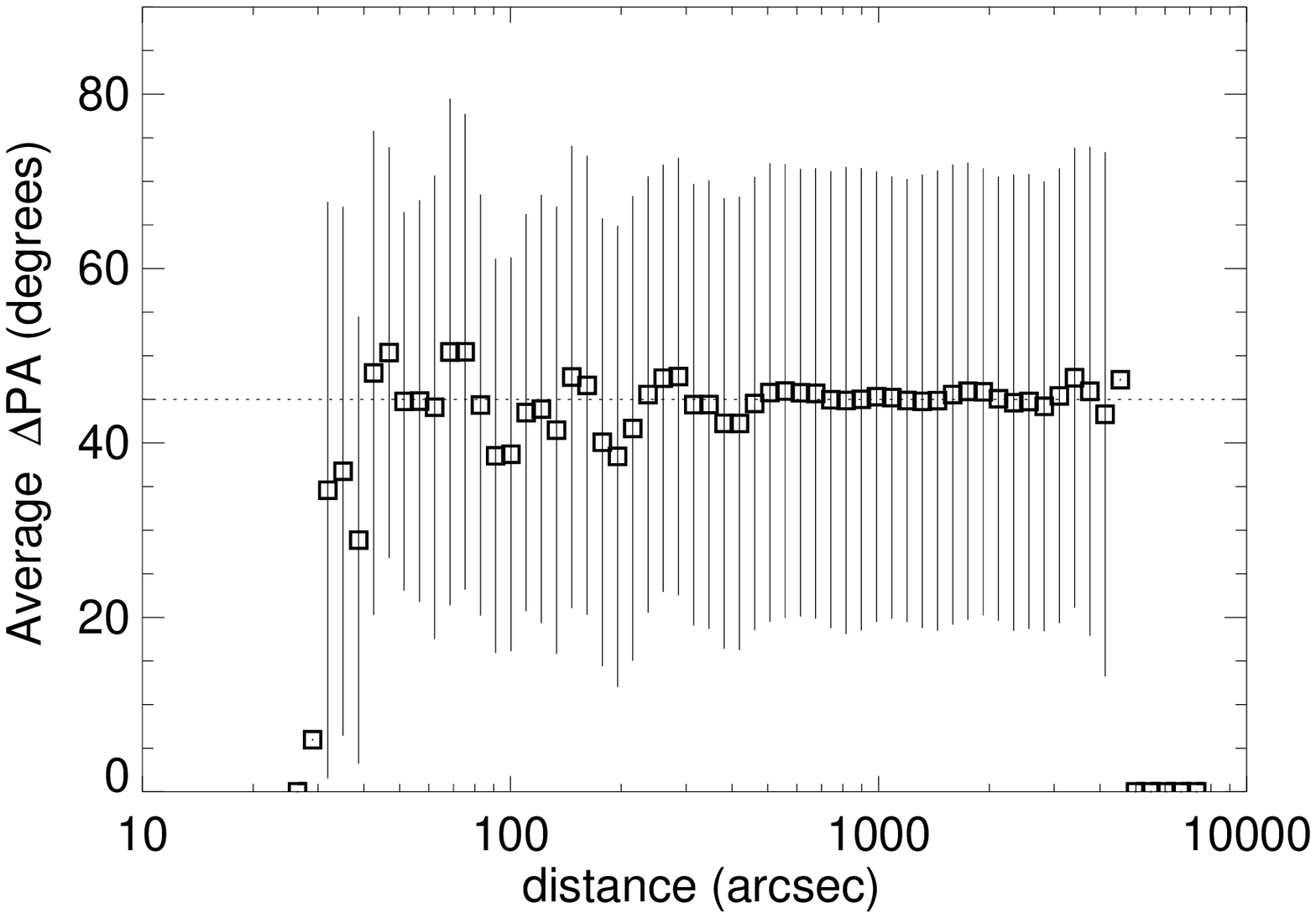,width=8.5cm,angle=0}}
\caption{The relative orientation (difference in position angles) of core
  major axes with respect to their neighbours: average relative orientation
  with respect to the $n$th neighbour (upper panel), and average relative
  orientation  of cores within a certain (projected) distance to the core.
  The smallest cores (with major axes determined to be smaller than 28\arcsec)
  have been omitted. }
\label{relorient}
\end{figure}

To further check for the presence of local (or global) alignment of core
orientations, we have analysed the absolute difference in position angles 
($\Delta$PA) that every
core includes with its neighbours. Fig.\ \ref{relorient} (upper panel) shows
the average position angle difference of all cores with respect to their
$n$'th nearest neighbour. The large scatter in $\Delta$PA around 45$^\circ$
indicates that
the large majority of cores is randomly oriented with respect to their 
neighbours. At most a tiny tendency towards lower average $\Delta$PA
is seen for the few very next neighbours; the still large scatter indicates
that this can only be due to very few cases showing some degree of alignment,
while the great majority is randomly oriented.

Fig.\ \ref{relorient} (lower panel) shows the average $\Delta$PA as a function
of the projected distance from every core. Again, at most a very small
trend towards smaller average $\Delta$PA at the smallest separations can
be seen. The large scatter again indicates that this trend
is caused by only a few core pairs, with most core pairs including random
$\Delta$PA.

Thus it appears that there is virtually no tendency for alignment of
core orientations, supporting the idea of core formation as a result of
unordered, turbulent motions. In particular, supposing that the orientation
of the core major axis is perpendicular to the direction of preferred 
contraction, this contradicts the idea of ordered magnetic fields as producing
agent of aligned protostellar disk/jet systems.

Two (counteracting) effects should be kept in mind in the interpretation of
this result: residual scanning effects might locally push measured core
orientations towards a similar position angle. On the other hand, we only
can analyse the projected core distribution; a core seen to be the nearest 
neighbour in this 2-D projection might in fact be at much larger distance
in 3-D, which will tend to dilute any local alignment, thus in fact alignment
might be more prevalent than seen here.

\end{document}